\documentclass[a4paper,11pt]{article}
\usepackage[text={17.1cm,24.6cm},centering]{geometry}
\usepackage[utf8]{inputenc}
\usepackage{physics}
\usepackage{amsmath,amssymb,mathtools}
\usepackage{subfigure}
\usepackage{pst-node}
\usepackage{tikz-cd} 
\usepackage{amssymb}
\usepackage{graphicx}
\usepackage{authblk}
\usepackage{cite}
\usepackage[colorlinks=true,linkcolor=blue, citecolor=red, urlcolor=blue, bookmarks]{hyperref}

\title{Symmetry resolved entanglement entropy of excited states in a CFT}
\author[1]{Luca Capizzi\footnote{email: lcapizzi@sissa.it}}
\author[2]{Paola Ruggiero}
\author[1,3]{Pasquale Calabrese}
\affil[1]{SISSA and INFN, Via Bonomea 265, 34136 Trieste, Italy}
\affil[2]{DQMP, University of Geneva, 24 Quai Ernest-Ansermet, CH-1211 Geneva, Switzerland}
\affil[3]{International Centre for Theoretical Physics (ICTP), Strada Costiera 11, 34151 Trieste, Italy}
\date{}                     
\begin{document}
\maketitle

\begin{abstract}
We report a throughout analysis of the entanglement entropies related to different symmetry sectors in the low-lying primary excited states of a 
conformal field theory (CFT) with an internal $U(1)$ symmetry. Our findings extend  recent results for the ground state.
We derive a general expression for the charged moments, i.e. the generalised cumulant generating function, which can be written in terms of correlation 
functions of the operator that define the state through the CFT operator-state correspondence. 
We provide explicit analytic computations for the compact boson CFT (aka Luttinger liquid)  for the vertex and derivative excitations.
The Fourier transform of the charged moments gives the desired symmetry resolved entropies. 
At the leading order, they satisfy entanglement equipartition, as in the ground state, but we find, within CFT, subleading terms that break it. 
Our analytical findings are checked against free fermions calculations on a lattice, finding excellent agreement.
As a byproduct, we have exact results for the full counting statistics of the $U(1)$ charge in the considered excited states.
\end{abstract}

\baselineskip 18pt
\thispagestyle{empty}
\newpage

\tableofcontents

\section{Introduction}

Symmetries play a central role in all fields of modern physics from the phenomenology of the standard model (and beyond) to the theory of phase transition, 
from string theory to solid-state passing through nuclei and molecules: it is impossible to overestimate their importance as a guiding principle in our 
current understanding of the physical world. 
One fundamental aspect of symmetries is Noether theorem: any continuous symmetry corresponds to a conservation law of local degrees of freedom,  
in both classical and quantum physics. 
For example, the rotational symmetry of an interacting spin chain implies that some components of total spin is preserved during the time evolution.
Hence, the presence of symmetries is a huge constraint for the dynamics of many-body systems and it is therefore important to understand the properties of physical 
states under the related symmetry transformations. 
In particular, it is rather natural to wonder about the implications of symmetries for the entanglement content of physical  states of extended
quantum systems with internal symmetries,  a subject that, very surprisingly, got attention only in very recent times. 

The most useful and successful way of characterising the bipartite entanglement in a many-body quantum system is through the R\'enyi  entropies 
of the reduced density matrix $\rho_A$ of  a given subsystem $A$.
These are defined as 
\begin{equation} \label{defSn}
	S_n \equiv \frac{1}{1-n} \log \textrm{tr} \left( \rho_A^n \right),
\end{equation}
where the index $n$ is an arbitrary (positive) real parameter, but in many cases, as we shall see, it is useful to think to it as an integer. 
Among the R\'enyi  entropies, the von Neumann one
\begin{equation}
	S \equiv  \lim_{n \to 1} S_n = S_1= - \textrm{tr} ( \rho_A \log \rho_A ),
\label{Svn}
\end{equation}
has a special place.
Nowadays, the entanglement entropies are standard tools in the study and analysis of many-body quantum systems:
they are largely employed to detect quantum phase transitions \cite{hlw-94,vlrk-03,cc-04,cc-09}; 
they proved to give a deeper understanding of topological features of condensed matter systems such as quantum Hall states \cite{lihaldane}; 
and much more (see \cite{amico-2008,calabrese-2009,eisert-2010,rev-lafl} as reviews of applications).
In the context of quantum field theory (QFT), they are typically calculated and analysed in a \emph{replica approach} \cite{cc-04,cc-09}; 
in the special case of $(1+1)$-dimensional conformal field theories (CFT) explicit analytic results can be obtained  in 
many different situations and states.

For quantum systems with internal symmetries, one can identify the contributions to the entanglement coming from each symmetry sector through the \emph{symmetry resolved} 
entanglement (see Section \ref{sec:sela} for explicit definitions).
Nonetheless, computing them analytically in a many-body quantum system remained a hard task until recently, when a new theoretical framework has been 
introduced in Refs.~\cite{Gsela,GS-neg}. 
The new main insight of these works is to relate the symmetry resolved quantities to the path integral over a Riemann surface, where twisted boundary conditions are imposed 
along the branch cuts (see Section \ref{sec:sela} for further details), in turn easily computed using a simple modification of the known replica tricks.
After these initial works, there has been a large effort in characterising the symmetry resolved entanglement in the ground state of many-body systems. 
In fact, for one-dimensional (1D) systems, several results are known for CFTs \cite{Gsela, equi-sierra}, 
free gapped and gapless systems of bosons and fermions \cite{sara-gapped, Ric, Gold}, 
and integrable spin chains \cite{sara-gapped,sara-fcs,lr-14}.
Very recently, few results appeared also in higher dimensions \cite{Gold, ryu-symmres,mrc-20}. 
The out-of-equilibrium behaviour of symmetry resolved entanglement after a local quantum quench has been investigated as well \cite{fg-19}. 
Finally,  the relevance of the symmetry resolution of the entanglement in the non-equilibrium dynamics of disordered systems has been underlined in \cite{exp-lukin},
even from the experimental point of view.

The main goal of this work is to generalise the ground-state CFT approach for the symmetry resolved entanglement \cite{Gsela, equi-sierra} to excited states. 
The total von Neumann and R\'enyi entropies in excited states were first considered in \cite{excSierra, excSierra2}, where 
the replica trick for the ground state \cite{cc-04} was generalised to treat these more complicated states.
By combining the results of \cite{Gsela} with the ones of \cite{excSierra, excSierra2}, we provide full analytical results for the symmetry resolved entanglement entropies.

The paper is organised as follows.
In Section \ref{sec:sierra}, we briefly recall the results of \cite{excSierra, excSierra2} for the excited states' entanglement. 
In Section \ref{sec:sela} we provide all the definitions concerning symmetry resolved entanglement measures  and summarise  the known 
results in the ground state. 
Our main findings for the symmetry resolved entanglement  in excited states are reported in Sections \ref{new} and \ref{ana_results}: 
in the former we report the general treatment in an arbitrary CFT and in the latter we specialise to  
the massless compact boson; numerical checks for free fermions on the lattice are given in Section~\ref{num_results}
together with some details about their implementation. 
We conclude in Section \ref{sec:final} with some discussions and speculating about future directions.

\section{Entanglement of excited states in CFT} \label{sec:sierra}

Here, we briefly summarise the replica approach  to the entanglement entropies of excited states in a $(1+1)$-dimensional CFT as developed in Refs. \cite{excSierra, excSierra2}. 
Let $L$ be the total length of a periodic 1D system, $A$ a subsystem consisting of a segment of length $\ell$ (say $A=[u,v]$ with $v-u=\ell$) and $B$ its complement. 
In the path integral approach,  the ratio between $\text{tr}(\rho_A^n)$ (with $n$ integer) in a given excited state and 
the one in the ground state is written as a ratio of correlation functions on specific Riemann surfaces;
such ratios turn out to be universal functions of  $x \equiv {\ell}/{L}$. 
From there, it is possible to extract the excess of R\'enyi entropy that the excited state has with respect to the ground state in the bipartition $A\cup B$. 
Note that, one has to focus on a finite value of $L$ (instead of an infinite system) in order to observe some excess of entropy: 
for low-lying excited states, this excess vanishes in the thermodynamic limit.

The starting point is that an excited state $\ket{\Upsilon}$ may be written as the insertion of a local operator $\Upsilon(x,\tau)$  at past infinite $\tau = -\infty$ as
\begin{equation}
\ket{\Upsilon} \sim \underset{\tau \rightarrow -\infty}{\lim} \Upsilon(x,\tau)\ket{0},
\label{UPstate}
\end{equation}
where $\ket{0}$ is the ground state of the CFT. This mapping is known as state-operator correspondence (see, e.g., \cite{pagialle} for details) 
and applies to any state of the Hilbert state of the CFT. 
The corresponding path-integral representation of the density matrix $\rho=|\Upsilon\rangle\langle\Upsilon|$ presents two insertions of $\Upsilon$ at 
$z=x+i\tau =\pm i\infty$.
The world sheet is an infinite cylinder of circumference $L$.

In what follows we omit the index $A$ of the density matrix $\rho_A$, denoting by $\rho_\Upsilon \equiv \text{tr}_{B}(\ket{\Upsilon}\bra{\Upsilon})$ the reduced density 
matrix associated with the state $\ket{\Upsilon}$; 
we denote the reduced density matrix of the ground state by $\rho_{\mathbb{I}}$,  thinking to the ground state as the one corresponding to 
the identity operator $\mathbb{I}$.
Next, following Refs. \cite{excSierra, excSierra2}, we define the ratio
\begin{equation}
F_{\Upsilon}^{(n)}(x) \equiv \frac{\text{tr}(\rho_\Upsilon^n)}{\text{tr}(\rho_{\mathbb{I}}^n)},
\label{FUPdef}
\end{equation}
so to have a universal quantity which is neither UV-divergent nor depends on microscopic scales (as it is instead the case for $\text{tr}(\rho_{\Upsilon}^n)$). 
For an arbitrary operator $\Upsilon$, $\text{tr}(\rho_{\Upsilon}^n)$ may be be obtained by sewing cyclically (along the interval $[u,v]$) 
$n$ of the above cylinders defining the reduced density matrix $\rho_{\Upsilon}$. 
In this way, one arrives at a $2n$-point function of $\Upsilon$ on a $n$-sheeted Riemann surface $\mathcal{R}_n$.
Keeping track of the  correct normalisation of $\rho_{\Upsilon}$, one straightforwardly obtains \cite{excSierra, excSierra2}
\begin{equation}
F_{\Upsilon}^{(n)}(x) =  \frac{\left\langle \displaystyle \prod_{k=0}^{n-1} \Upsilon(z_k^-) \Upsilon^\dagger(z_k^+) \right\rangle_{\mathcal{R}_n}    }{\langle \Upsilon(z_0^-) \Upsilon^\dagger(z_0^+) \rangle_{\mathcal{R}_1}^n },
\label{FU}
\end{equation}
where $z_k^{\mp}$ corresponds to the points at past/future infinite respectively of the $k$-th copy of the system ($k=0,...,n-1$) in $\mathcal{R}_n$ 
(${\cal R}_1$ is just the cylinder). 
The normalisation  factor of the field $\Upsilon$ does not matter because it cancels out in the ratio \eqref{FU}; 
moreover, $F^{(1)}_\Upsilon(x)=1$ as it should be because of the normalisation of the involved density matrices.

Through the conformal mapping \cite{excSierra2}
\begin{equation}
w(z) = -i\log \left(-\frac{\sin\frac{\pi(z-u)}{L}}{ \sin\frac{\pi(z-v)}{L}}\right)^{1/n},
\label{transf}
\end{equation}
the Riemann surface $\mathcal{R}_n$ is transformed into a single cylinder. 
At this point, exploiting the transformation of the field $\Upsilon$ under a conformal mapping, one relates the ratio \eqref{FU} to the correlation functions of $\Upsilon$ on the plane.  
For this reason, afterward we focus on those low-lying states described by primary operators of the CFT. 
This assumption is not fundamental but simplifies the treatment due to the simple transformation law of such operators, i.e. 
\begin{equation}
\Upsilon(w, \bar{w}) = \left( \frac{dz}{dw}\right)^h  \left( \frac{d\bar{z}}{d\bar{w}}\right)^{\bar{h}} \Upsilon(z,\bar{z}),
\end{equation}
with $(h,\bar{h})$ the conformal weights of $\Upsilon$. Hence, for primary operators, one can easily express $F^{(n)}_{\Upsilon}(x)$ in terms of correlation functions over the cylinder. 
The final result reads \cite{excSierra2}
\begin{equation}
F_\Upsilon^{(n)}(x) =  n^{-2n(h+\bar{h})}  \frac{\langle\prod_k \Upsilon(w^-_k)\Upsilon^\dagger(w^+_k)\rangle_{\rm cyl}}{\langle\Upsilon(w^-_0)\Upsilon^\dagger(w^+_0)\rangle^n_{\rm cyl}},
\label{FU2}
\end{equation}
where $w_k^{\pm}$ are the points corresponding to $z_k^{\pm}$ through the map $w(z)$, i.e. 
\begin{equation}
 w^-_k = \frac{\pi(1+x)+2\pi k}{n},\qquad
 w^+_k = \frac{\pi(1-x)+2\pi k}{n}\,, \qquad {\rm with}\;  {k=0,...,n-1}.
\end{equation}
Translational invariance  ($w\rightarrow w+r$  with $r\in \mathbb{R}$) and parity ($w \rightarrow-w$) of the cylinder imply 
\begin{equation}
F_\Upsilon^{(n)}(x) = \overline{F_\Upsilon^{(n)}(x)} = F_{\Upsilon^\dagger}^{(n)}(x) = F^{(n)}_{\Upsilon}(1-x),
\end{equation}
which, among the other things, guarantee the symmetry $\ell \rightarrow L-\ell$ of the R\'enyi entropies.

\subsection{The Luttinger liquid CFT} \label{subsec:LL}

In the following, we will explicitly work out the symmetry resolved entropies of the Luttinger liquid or equivalently (via bosonisation) 
of a free massless compact boson. The compactification radius of the boson is related to the  Luttinger parameter $K$.
The Luttinger liquid's universality class describes a large number of critical one-dimensional models including  free and interacting
spin-chains, quantum gases, fermionic hopping models, etc. (see e.g. \cite{giam-b}).
The central charge is $c=1$.
Denoting by $\varphi$ a real bosonic field, the euclidean action (in bosonic form) is
\begin{equation}
{\cal S}_E[\varphi] = \frac{1}{8\pi K}\int d\tau dx \ \partial_\mu \varphi \partial^\mu \varphi.
\label{S_bose}
\end{equation}
The field can be decomposed in holomorphic and antiholomorphic components, $\varphi(z,\bar{z}) = \phi(z)+\bar{\phi}(\bar{z})$.
As examples, the set of primary fields of the theory include the holomorphic vertex operators 
\begin{equation}
V_{\beta}(z) = e^{i\beta \phi(z)},
\end{equation}
and  the derivative operator
\begin{equation}
i\partial \phi(z).
\end{equation}
The scaling functions \eqref{FUPdef} of the moments of $\rho_\Upsilon$ for the excited states generated by the insertion of these primary operators 
as in Eq. \eqref{UPstate} have been 
obtained in \cite{excSierra2}, following the procedure outlined in the previous subsection. 
The final result for the vertex operator is $F^{(n)}_{V_\beta}(x)=1$, implying that all R\'enyi entropies of these excited states are the same as in the ground state. 
For the derivative operator, $F^{(n)}_{i\partial \phi}(x)$ is instead nontrivial and can be written as a $2n\times 2n$ determinant \cite{excSierra2};  
its analytical continuation has been obtained in Refs. \cite{det,elc-13}  and reads
 \begin{equation}
F^{(n)}_{i\partial \phi}(x) =  \left[ \left( \frac{2\sin(\pi x)}{n} \right)^n\frac{\Gamma(\frac{1+n+n \csc(\pi x)}{2})}{\Gamma(\frac{1-n+n \csc(\pi x)}{2})} \right]^2.
\label{Fder}
\end{equation}
Other primary states are the antiholomorphic versions of these operators (and combinations thereof), to which similar results apply.
Also some results for non-primary operators and boundary theories are known, see e.g. \cite{palmai,top-16,cghw-15}.

\section{Symmetry resolved entanglement} \label{sec:sela}

We now consider a quantum system with an internal $U(1)$ symmetry and a bipartition in two spatial subsystems, $A$ and $B$.
Moreover, we assume that the quantum state with density matrix $\rho$ lies in a representation of the symmetry. 
For instance, if $Q$ is the generator of the $U(1)$ symmetry, we require that the state is an eigenvector of $Q$ with eigenvalue which identifies the underlying representation. 
Although $Q$ does not fluctuate, the presence of quantum correlations between $A$ and $B$ usually reflects  in the fluctuations of 
$Q_A = \text{tr}_B(Q)$ and $Q_B =  \text{tr}_A(Q)$. 
Since the statistical properties of $Q_A$ are encoded in the reduced density matrix $\rho_A = \text{tr}_B(\rho)$, one may study the decomposition of the spectrum of $\rho_A$ in the different eigenspaces of $Q_A$, which are the sectors of the $U(1)$ symmetry of the subsystem $A$.

For a $U(1)$ charge (more generally for any additive charge), the commutator $[\rho,Q]=0$ implies $[\rho_A,Q_A]=0$, by simply tracing out the subsystem $B$. 
Hence, $\rho_A$ has a block diagonal structure with each block corresponding to an eigenvalue $q$ of $Q_A$. 
One can thus relate a conditioned density matrix $\rho_A(q)$ to any eigenvalue $q$; $\rho_A(q)$ is obtained by projecting $\rho_A$ onto the eigenspace of $Q_A$
with fixed $q$, as induced by the projector $\Pi_q$, i.e. 
\begin{equation} \label{rhoAq}
\rho_A(q) \equiv \frac{\rho_A \Pi_q}{\text{tr}(\rho_A \Pi_q)}.
\end{equation} 
The denominator is introduced to force the normalisation ${\rm tr} \rho_A(q)=1$.

Consequently, we can  define  {\it symmetry-resolved} entanglement entropy $S(q)$ and R\'enyi entropies $S_n(q)$ for each sector where $Q_A = q$; 
this is the amount of entanglement shared by $A$ and $B$ in each symmetry sector.
The symmetry resolved R\'enyi entropies are
\begin{equation} \label{SnqSq}
S_n(q) \equiv \frac{1}{1-n} \log \trace(\rho_A(q)^n) = \frac{1}{1-n} \log \frac{ \trace(\rho_A^n \Pi_q) }{\trace(\rho_A\Pi_q)^n  },
\end{equation}
with von Neumann limit
\begin{equation} \label{Sq}
S(q) = \underset{n\rightarrow 1}{\lim} S_n(q)  = -\frac{ \trace(\rho_A \log \rho_A   \Pi_q)}{\trace(\rho_A \Pi_q)} + \log \trace(\rho_A \Pi_q)=-{\rm tr}[ \rho_A(q) \log \rho_A(q)].
\end{equation}
Notice that, in this language, the probability distribution of the charge is 
\begin{equation}
p(q) = \trace(\rho_A\Pi_q).
\end{equation}
Taking the average of $S(q)$ with respect to the charge $q$ (i.e. multiplying both side of Eq. \eqref{Sq} by $p(q)$ and summing over $q$), one obtains
\begin{equation} \label{2terms}
S= \langle S(q) \rangle_p - \sum_q p(q)\log p(q),
\end{equation}
where we introduced $\langle S(q) \rangle_p \equiv \sum_q p(q)S(q)$ and $S$ is the total entropy in Eq. \eqref{Svn}. 
Equation \eqref{2terms} shows that the total entropy is larger than the averaged symmetry resolved entropy (equivalently their weighted sum)  
and the difference is the Shannon entropy related to the probability distribution of $Q_A$, i.e. $- \sum_q p(q)\log p(q)= \langle -\log p(q)\rangle_p$.
The two terms in \eqref{2terms} are usually referred to as \emph{configurational} and \emph{fluctuation} entanglement, respectively \cite{exp-lukin}.
Note that the configurational entropy is also related to the operationally accessible entanglement entropy \cite{wv-03,bhd-18,bcd-19}. 

In general, the calculation of the symmetry resolved entropies requires the knowledge of the spectrum of $\rho_A$ and its resolution in $Q_A$. 
However, this is a rather difficult task, especially for an analytic derivation.  
The main idea put forward in Ref.~\cite{Gsela} (see also \cite{equi-sierra}) is that the same result can be  achieved by 
focusing on the computation of the {\it charged moments}
\begin{equation}
Z_n(\alpha)\equiv\trace(\rho_A^n e^{i\alpha Q_A}).
\end{equation}
In fact,  the Fourier transform of the charged moment $Z_n(\alpha)$ with respect to $\alpha$ gives  $\trace(\Pi_q \rho_A^n)$ 
and thus it is thus directly related to $S_n (q)$ through \eqref{SnqSq}. Similar charged moments have been already considered in the context of free field theories \cite{CFH,d-16,ch-rev}, in holographic settings \cite{matsuura,cnn-16}, as well as in the study of entanglement in mixed states \cite{ssr-17,shapourian-19}.

Eq. \eqref{2terms} is valid only for the von Neumann entropy and it is not possible to write down an analogue formula for the R\'enyi in terms of the probability $p(q)$.
This no-go result could be at first disappointing, but  it can be circumvented by defining 
\begin{equation}
p_n(q)\equiv \frac{\trace(\rho_A^n \Pi_q)}{\text{tr}(\rho_A^n)}.
\label{p_n}
\end{equation}
While for $n=1$, $p_1(q)$ is just $p(q)$, the physical probability distribution of the charge $Q_A$, for $n \neq 1$ there is no direct meaning of the 
probabilities $p_n(q)$ for $n\neq1$, although they are normalised as $\sum_q p_n(q)=1$. 
However, these probabilities $p_n(q)$ are useful since they allow us to write $S_n(q)$ as 
\begin{equation}
S_n(q) = \frac{1}{1-n} \log \frac{ \trace(\rho_A^n \Pi_q) }{\trace(\rho_A\Pi_q)^n  }= 
\frac{1}{1-n} \log \frac{ \trace(\rho_A^n \Pi_q) }{\text{tr}(\rho_A^n)}\frac{\text{tr}(\rho_A^n)}{\trace(\rho_A\Pi_q)^n  }=
S_n + \frac{1}{1-n}\log \frac{p_n(q)}{p(q)^n},
\label{eq11}
\end{equation}
in which the entire $q$-dependence is in the second term being $S_n$  the total R\'enyi entropy of Eq. \eqref{defSn}.
The limit for $n\to1$ of the above is
\begin{equation}
S(q) = S - \frac{\partial_n p_n(q)|_{n=1}}{p(q)}+ \log p(q).
\label{eq1}
\end{equation}
The average of \eqref{eq1} over $p(q)$ gives back Eq. \eqref{2terms} after using $\sum_q \partial_n p_n(q)|_{n=1}=0$ (which follows from the derivative wrt $n$ of the 
normalisation of $p_n(q)$, i.e.  $\sum_q p_n(q)=1$).
It is also useful to average Eq. \eqref{eq11} over $p(q)$ to obtain the two equivalent forms 
\begin{multline}
S_n= \langle S_n(q) \rangle_p -\frac1{1-n} \sum_q p(q) \log \frac{p_n(q)}{p(q)^n}=\\= 
\langle S_n(q) \rangle_p - \sum_q p(q) \log p(q) -\frac1{1-n} \sum_q p(q) \log \frac{p_n(q)}{p(q)}.
\label{snpn}
\end{multline}
In the last expression the first term is  the averaged symmetry resolved R\'enyi entropy, i.e. a configurational R\'enyi entropy analogous to the von Neumann one 
in Eq. \eqref{2terms}; the second term is just the fluctuation von Neumann entropy, identical to the one in Eq. \eqref{2terms}; 
the third term is instead new and it is the only one related to the probability $p_n(q)$ which makes not possible to write $S_n$ only in terms of $p(q)$. 
A similar expression can be also written as average over $p_n(q)$ instead of $p(q)$ (in the first line in Eq. \eqref{snpn} only the probability for the average changes). 
Eq. \eqref{snpn} is different from the R\'enyi fluctuation  entropy $S_F^{(n)}\equiv (\log \sum_q (p(q))^n)/(1-n)$ defined, e.g., in Ref. \cite{kusf-20}.

The Fourier transform of the generalised probability $p_n(q)$ in Eq. \eqref{p_n} is
\begin{equation}
p_n(\alpha) \equiv \frac{\text{tr}(\rho_A^ne^{i\alpha Q_A})}{\text{tr}(\rho_A^n)} ,
\label{p_al}
\end{equation}
and it is a (normalised) moment generating function for  $p_n(q)$.

\subsection{Replica method and CFT}

The moments $\text{tr}(\rho_A^n)$ have a  geometrical interpretation for any $(1+1)$ QFT in terms of a partition function over a Riemann surface $\mathcal{R}_n$;
such geometrical approach leads to universal results for the ground states of $(1+1)$ CFT's. 
Following Ref. \cite{Gsela}, we give a geometrical meaning also to $\text{tr}(\rho_A^ne^{i\alpha Q_A})$.
To this aim, let us introduce a local operator $\mathcal{V}_\alpha(x,\tau=0)$ which implements the $U(1)$ symmetry, acting as a phase shift of $e^{i\alpha}$ in 
the spacial subregion $[x,+\infty)$ (see \cite{Gsela,Ric} for details). If $A$ is the segment $[u,v]$, one can thus identify
\begin{equation}
e^{i\alpha Q_A} = \mathcal{V}_\alpha(u, 0)\mathcal{V}_{-\alpha}(v, 0).
\label{Vdef}
\end{equation}
When $\rho$ is the ground state of a QFT, $\text{tr}(\rho_A^ne^{i\alpha Q_A})$ can be seen as a partition function over a Riemann surface with twisted boundary conditions (introducing a phase factor $e^{i\alpha}$ between the first and the last sheet along $A$) or, equivalently, as a correlation function $\langle \mathcal{V}_\alpha(u, 0)\mathcal{V}_{-\alpha}(v,0)\rangle_{\mathcal{R}_n}$ over the Riemann surface $\mathcal{R}_n$ with periodic boundary conditions.

In the ground state of a CFT, if one specialises to an infinitely extended system and when ${\cal V}_\alpha$ is a primary operator, 
the scaling of $\text{tr}(\rho_A^ne^{i\alpha Q_A})$ is determined by the value of $c$, the central charge of the underlying theory, and $(h_{\alpha}, \bar{h}_{{\alpha}})$, 
the conformal weights of ${\cal V}_\alpha$, through \cite{Gsela}
\begin{equation}
\text{tr}(\rho_A^ne^{i\alpha Q_A}) \sim \ell^{-\frac{c}{6}(n-\frac{1}{n})-\frac{h_{\alpha}+h_{{-\alpha}} + \bar{h}_{\alpha} + \bar{h}_{{-\alpha}}   }{n}},
\label{Grstate}
\end{equation}
having denoted as $\ell \equiv v-u$ the length of the region $A$. Similar conclusions applies to finite systems of length $L$ through the replacement \cite{cc-04}
\begin{equation}
\ell\ \rightarrow \frac{L}{\pi}\sin\frac{\pi \ell}{L},
\label{finite}
\end{equation}
following from the conformal map from the plane to the cylinder. 

The previous results apply to a generic $U(1)$ charge. Hereafter, we specialise to the $U(1)$ symmetry of the Luttinger liquid or compact boson 
defined by the action \eqref{S_bose}. In this case, the conserved current is proportional to $\partial_x \varphi$ and hence 
the charge operator in the interval $A$ is 
\begin{equation}
Q_A = \frac{1}{2\pi}\int_A dx \ \partial_x \varphi = \frac{1}{2\pi}(\varphi(v)-\varphi(u)).
\end{equation}
Hence, by simple inspection of Eq. \eqref{Vdef}, the local operator ${\cal V}_\alpha$ is implemented by the  vertex operator
\begin{equation}
\mathcal{V}_\alpha =V_{\frac{\alpha}{2\pi}}\equiv e^{i\frac{\alpha}{2\pi} \varphi},
\label{vert}
\end{equation}
with weights $(h_{\alpha}, \bar{h}_{{\alpha}}) = \left(\frac{K}{2}(\frac{\alpha}{2\pi})^2,\frac{K}{2}(\frac{\alpha}{2\pi})^2 \right)$ (it contains both the holomorphic 
and the antiholomorphic sector). 
In this case, the Fourier transform of $p_n(q)$ (cf. $p_n(\alpha)$ in Eq. \eqref{p_al}), is gaussian
\begin{equation}
p_n(\alpha)= \frac{\text{tr}(\rho_A^ne^{i\alpha Q_A})}{\text{tr}(\rho_A^n)} \sim \ell^{-\frac{\alpha^2}{2}\frac{K}{n \pi^2}},
\label{gaussGS}
\end{equation}
and, as a consequence, also $p_n(q)$ itself is gaussian. Therefore, in this CFT, $p_n (q)$ is fully characterised by its variance.

Let us now conclude this section by discussing  the consequences of our findings for 
a microscopical model that, at large scale, displays conformal invariance and it is in the Luttinger liquid universality class
(such as, for example, the XXZ spin chain, the one-dimensional Bose gas and many more). 
For these models, conformal invariance fixes only the universal part of the scaling form of the distribution $p_n(q)$ but it does not predict other non-universal contributions 
which may play some role also in the limit $\ell \rightarrow \infty$. 
For concreteness, let us focus on the variance $\langle \Delta q^2 \rangle_n $, calculated as average over the distribution $p_n(q)$,
which for large $\ell$ scales as 
\begin{equation}
\langle \Delta q^2\rangle_n\equiv \sum_q p_n(q) (q-\bar q)^2
 = \frac{K}{n \pi^2} \log \ell + b_n+o(1),
\label{varianze}
\end{equation}
as follows straightforwardly from Eq. \eqref{gaussGS}.
In Eq. \eqref{varianze}, the multiplicative factor of the logarithm is fixed by CFT and it is universal;   
instead the additive constant $b_n$ depends on the details of the model (it is fixed by the non-universal amplitude not specified in Eq. \eqref{gaussGS}). 
For example, in the XX chain (whose underlying CFT is a Luttinger liquid with $K=1$) the exact value of $b_n$ has been derived exploiting 
the Fisher-Hartwig conjecture \cite{Ric}. 
The value $\bar q$ also is not fixed by CFT and requires a microscopic computation. 
Note also that, for finite size systems, the replacement \eqref{finite} is equivalent to redefine $b_n$ as
\begin{equation}
b_n \rightarrow b_n +\frac{K}{n\pi^2}\log(\frac{1}{\pi x}\sin \pi x).
\end{equation}
Hence,  at order $O(1)$, the probability $p_n(q)$ is still gaussian, and so, in terms of the variance, we can write it as
\begin{equation} \label{pnqVar}
p_n(q) = \frac{1}{\sqrt{2\pi \langle \Delta q^2\rangle_n}}\exp(-\frac{\Delta q^2}{2\langle \Delta q^2\rangle_n}),
\end{equation}
where $\Delta q^2\equiv (q-\bar q)^2$.
Notice that in a lattice microscopical model (such as a spin chain), the integration over $\alpha$ does not run on the entire real axis but only on the interval
$\alpha\in[-\pi,\pi]$. However, because of the Gaussian form of $p_n(\alpha)$ with the variance \eqref{varianze}, this change of domain of integration 
only provides subleading corrections to \eqref{pnqVar}. 

The probability distribution \eqref{pnqVar} is all we need to determine the scaling for the symmetry resolved entanglement $S_n (q)$, which, 
 in the physical regime with $\Delta q$ of order 1, turns out to be
\begin{multline}
 S_n(q) = S_n -\frac{1}{2} \log \Big(\frac{2K}{\pi} \log \ell \Big)+ \frac{\log n}{2(1-n)}+\\ 
 \frac{1}{2(1-n)(\frac{K}{n\pi^2} \log \ell)}(b_1 - b_n) + \frac{\Delta q^2+\frac12}{2(n-1)(\frac{K}{n\pi^2} \log \ell)^2}\left(\frac{b_1}{n}-b_n \right) + o(\log \ell^{-2})=\\
 S_n -\frac{1}{2} \log \Big(\frac{2K}{\pi} \log \delta_n \ell \Big)+ \frac{\log n}{2(1-n)}+ 
 \Big(\Delta q^2+\frac12\Big)\frac{b_1-nb_n}{2n(n-1)(\frac{K}{n\pi^2} \log \ell)^2} + o(\log \ell^{-2}).
\label{Dom}
\end{multline}
Let us critically discuss this form. 
The leading term of $S_n(q)$ of order $\log \ell$ is equal to the total entropy $S_n$ (hence there are no contributions at the leading order from the second term in Eq.~\eqref{eq11}). 
The first subleading term behaves like $-\frac{1}{2}\log \log \ell + O(\ell^0)$ and also does not depend on $q$. 
The fact that the leading terms in $S_n(q)$ do not depend on $q$ has been dubbed \textit{equipartition of entanglement} \cite{equi-sierra}.
The other subleading terms can be written as a formal expansion in $(\log \ell)^{-1}$. 
The first one in $(\log \ell)^{-1}$ is independent of $q$ and hence can be conveniently absorbed as the non-universal scale  $\delta_n$ in the $\log \log \ell$ term, 
as we did in the last line of Eq. \eqref{Dom}, closely following Ref. \cite{Ric}. 
The first term breaking the equipartition of entanglement appears at order $\frac{\Delta q^2}{(\log \ell)^{2}}$ and its amplitude is nonuniversal. 
We mention that all non-universal constants in Eq. \eqref{Dom} have been exactly calculated in \cite{Ric} for the tight-binding model 
and they are important to correctly reproduce numerical results \cite{Ric}.

\section{Symmetry resolution of excited states} \label{new}

The main goal of this work is to obtain universal results for the symmetry resolved entanglement entropy in low lying excited states of CFT, in particular for Luttinger liquids. 
To this aim, we must combine the techniques for the symmetry resolved entanglement of Sec. \ref{sec:sela} with the CFT description of excited states in Sec. \ref{sec:sierra}.
We will combine these two concepts in this section, showing that, for all low lying excited states obtained by the action of a primary field as in Eq. \eqref{UPstate}, we can 
derive full analytic predictions for the universal scaling functions of the charged moments and from there draw conclusions for the 
symmetry resolved entanglement. 

We start by defining the universal function of interest for symmetry resolved entanglement of excited states.
For the total entanglement, the ratio of moments \eqref{FUPdef} is universal and can be computed in CFT without any input from the microscopical model \cite{excSierra}.
In the same spirit, we can define the $\alpha$-dependent ratio for the charged moments (we recall $x=\ell/L$)
 \begin{equation}
F_n(\alpha,x) \equiv \frac{\text{tr}(\rho^n_\Upsilon e^{i\alpha Q_A)}}{ \text{tr}(\rho^n_1 e^{i\alpha Q_A})},
\end{equation}
which is also universal and independent of any microscopic details. 
Notice that at $\alpha=0$, we have $F_n(0,x)=F_n(x)$. 
The latter observation suggests to define  another ratio
\begin{equation} \label{fU0}
f_n(\alpha,x) \equiv \frac{F_n(\alpha,x)}{F_n(0,x)}=
 \frac{\text{tr}(\rho^n_\Upsilon e^{i\alpha Q})\text{tr}\rho^n_1}{\text{tr}\rho^n_\Upsilon \text{tr}(\rho^n_1 e^{i\alpha Q})},
\end{equation}
which is also universal (it is the ratio of universal functions), but it has also the nice property $f_n(0,x) =1$ identically. 
Note that $f_n(\alpha,x)$ is nothing but the ratio of the generalised moment generating functions $p_n(\alpha)$ (cf. Eq. \eqref{p_al})
associated with the excited state $\ket{\Upsilon}$ and with the ground state $\ket{0}$ respectively
\begin{equation} \label{fnpns}
	f_n (\alpha,x) = \frac{p_n^{\Upsilon} (\alpha)}{p_n^{GS} (\alpha)}.
\end{equation}
Obviously, in these universal functions, all the non universal factors, e.g. coming from the variance of the $p_n$'s (cf. Eq. \eqref{varianze}), cancel.

\begin{figure}[t]
\begin{center}
\includegraphics[width = 0.45 \textwidth]{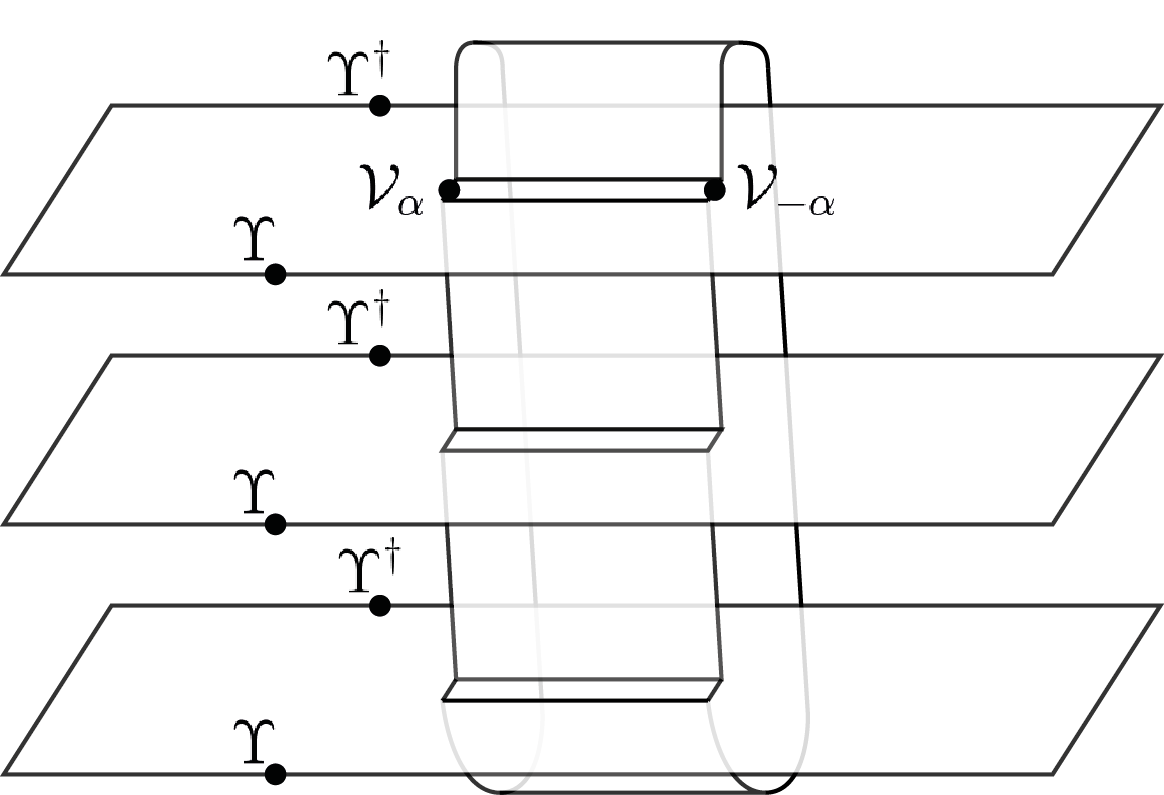}
\end{center}
\caption{Pictorial representation of  $\langle \mathcal{V}_\alpha(u_0)\mathcal{V}_{-\alpha}(v_0)\prod_k \Upsilon(z_k^-) \Upsilon^\dagger(z_k^+) \rangle_{\mathcal{R}_n}$, with $n=3$. This is the only new correlation arising in the calculation of the function $f_n (\alpha,x)$, cf. Eq.~\eqref{fU}, related to the (Fourier transform of the) symmetry 
resolved R\'enyi entropy in the excited state corresponding to the operator $\Upsilon$.}
\label{cartoon}
\end{figure}

The moments entering in the definition of $f_n(\alpha, x)$ in Eq. \eqref{fU0} may all be expressed as correlation functions of $\Upsilon$ and ${\cal V}_\alpha$ 
on the $n$-sheeted Riemann surface. 
Compared to the correlations defining $F_n(x)$ in Eq.~\eqref{FU} we only need to insert ${\cal V}_\alpha$ 
on an arbitrary sheet (we choose the 0-th one) at the branch points of the Riemann surface (i.e. $u_0,v_0$). 
Using the same conventions of Eq. \eqref{FU} for the insertions of $\Upsilon$ and $\Upsilon^\dagger$  (located at $\{z_k^{\mp}\}$, i.e. the past/future infinite respectively 
of the $k$-th copy).
we have
\begin{equation} \label{fU}
f_n(\alpha,x) = \frac{\displaystyle\left\langle \mathcal{V}_\alpha(u_0)\mathcal{V}_{-\alpha}(v_0)\prod_{k=1}^n \Upsilon(z_k^-) \Upsilon^\dagger(z_k^+) \right\rangle_{\mathcal{R}_n}    }{\displaystyle \langle \mathcal{V}_\alpha(u_0)\mathcal{V}_{-\alpha}(v_0) \rangle_{\mathcal{R}_n}\left\langle   \prod_{k=1}^n \Upsilon(z_k^-) \Upsilon^\dagger(z_k^+)   \right\rangle_{\mathcal{R}_n} }.
\end{equation}
The only correlation that has not yet been computed is the one in the numerator of \eqref{fU} which 
involves both $\mathcal{V}_\alpha$ and $\Upsilon$ and evidently is the most complicated one. 
A pictorial path-integral representation of this correlation is given in Fig.~\ref{cartoon}.
Notice that in Eq. \eqref{fU} there is no dependence on the normalisation of $\Upsilon$ and $\mathcal{V}_\alpha$, as it should.

All the correlation functions in \eqref{fU} are mapped by the conformal transformation \eqref{transf} to the cylinder.
Moreover, in this mapping, all powers of $(\frac{dz}{dw})$, coming from the transformation law of the primary operators, cancel out. 
Hence  $f_n(\alpha,x) $ may be  rewritten as  
\begin{equation}
f_n(\alpha,x) = \frac{\displaystyle\left\langle {\cal V}_\alpha(i\infty){\cal V}_{-\alpha}(i\infty)\prod_{k=1}^n \Upsilon(w_k^-) \Upsilon^\dagger(w_k^+) \right\rangle_{\rm cyl}    }{       \displaystyle  \langle \mathcal{V}_\alpha(i\infty)\mathcal{V}_{-\alpha}(-i\infty) \rangle_{\rm cyl}\left\langle   \prod_{k=1}^n \Upsilon(w_k^-) \Upsilon^\dagger(w_k^+)   \right\rangle_{\rm cyl} }.
\label{main}
\end{equation}

It is now useful to define the excess-cumulant generating function 
\begin{equation}
g_n (\alpha,x) \equiv \log f_n (\alpha,x)= \log \frac{p_n^{\Upsilon} (\alpha)}{p_n^{GS} (\alpha)}.
 \end{equation}
In fact, denoting with $\kappa_k^{\Upsilon}$ and $\kappa_k^{GS}$ the $k$-th cumulant in the state $\Upsilon$ and in the ground state respectively, 
we straightforwardly have
\begin{equation}
g_n (\alpha,x) =\sum_{k=1}^\infty \frac{\kappa_k^{\Upsilon}-\kappa_k^{GS}}{k!} (i\alpha)^k.
\label{cumuk}
\end{equation}
Hence, the first derivative of $g_n(\alpha,x)$ in ${\alpha=0}$  is the shift of the expectation value of $Q_A$ in going from the 
ground state to the excited state; the second derivative 
is the excess of the variance of $Q_A$, and so on for all other derivatives.
Hence, while all cumulants in general have non-universal contributions, the difference between a cumulant in the excited state and the same one in 
the ground state is {\it always universal}. 

Eq. \eqref{main} can be employed to calculate the charged moments of a primary excited state of an arbitrary CFT with a $U(1)$ symmetry (indeed it can 
be used for the resolution of an arbitrary symmetry, even non abelian, see e.g. Ref. \cite{Gsela}).
In the following section we specialise to the case of a Luttinger liquid CFT, introduced in Section~\ref{subsec:LL}.

\section{The Luttinger liquid CFT} 
\label{ana_results}

In the compact boson, there are two kinds of primary operators: the  vertex and the derivative operators.
In the following we work out the function $f_n (\alpha,x)$ for these two cases. 
We recall that for a Luttinger liquid the operator ${\cal V}_\alpha$ is a vertex operator (cf. Eq. \eqref{vert}) and so the calculation of $f_n(\alpha,x)$ 
just requires either the computation of multipoint correlation of vertices or of vertices and derivatives. 

\subsection{Vertex operator}

\begin{figure}[t]
\begin{center}
		\includegraphics[width=0.5\textwidth]{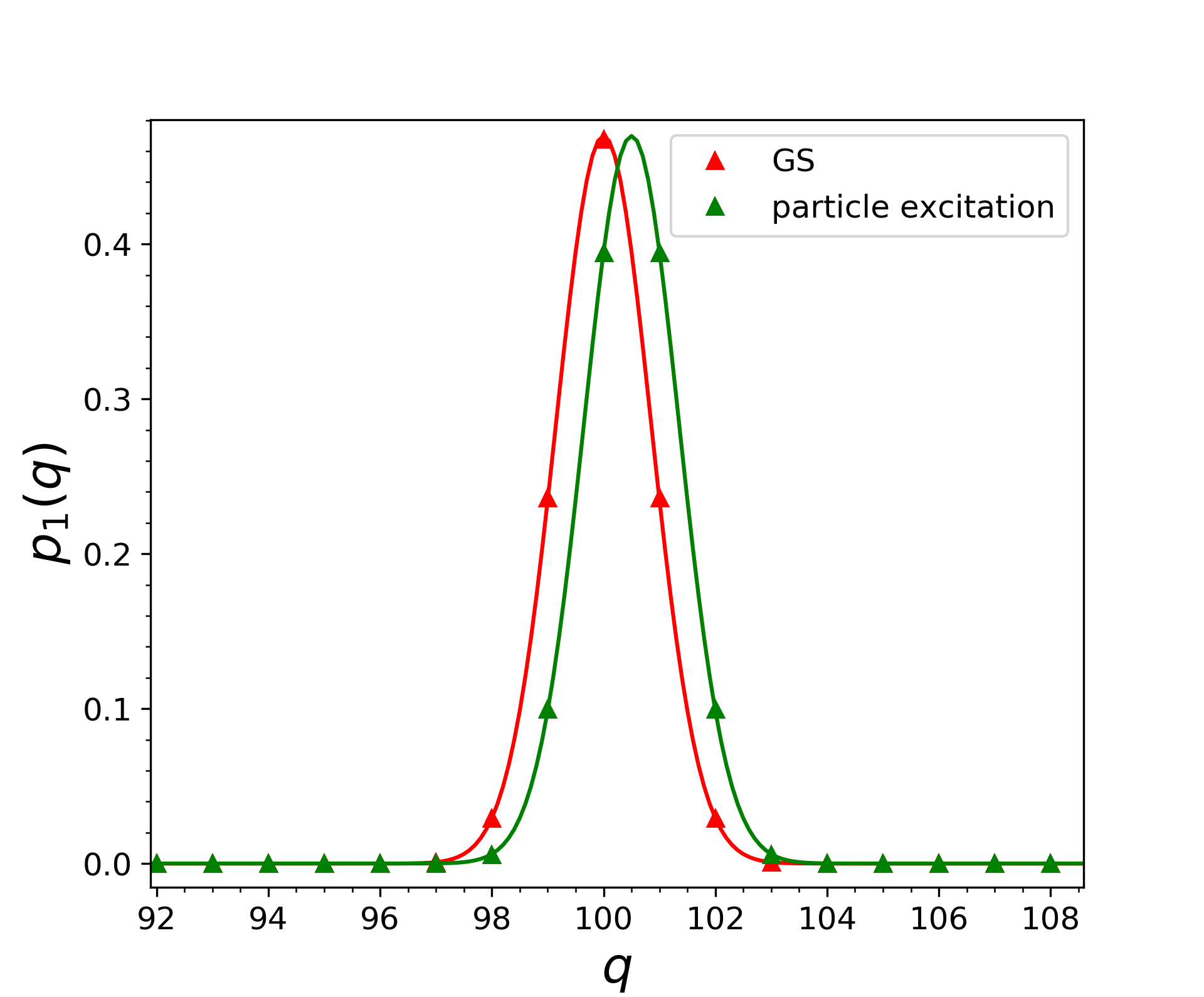}
\end{center}
\caption{Probability distribution of the number of particles in the ground state and in the excited state induced by the chiral vertex operator $e^{i \phi}$. 
The points are numerical data obtained from the half-filled XX chain (cf. Eq. \eqref{Hamiltonian}) for a system of length $L=400$ and a subsystem of size $\ell=L/2$. 
The excitation corresponds to a particle created over the Fermi sea in the lattice model. The continuous lines are gaussian probability distributions with variance given by $\langle \Delta q^2\rangle = \frac{1}{\pi^2}\log(\sin k_F \frac{L}{\pi}\sin \frac{\pi \ell}{L}) + \frac{1}{\pi^2}(1+\gamma_E+\log 2)$ 
(the non-universal $O(L^0)$ contribution can be found in \cite{Laf,Min}). 
The mean value of particles is $\langle q\rangle = 100.5$ in the excited state, while it is $\langle q\rangle = 100$ in the ground-state. 
}
	\label{prob1}
\end{figure}

The correlation functions of an arbitrary number of vertex operators $V_{\alpha_j}(z)$ are known by elementary methods \cite{pagialle} 
and on the cylinder may be written as 
 \begin{equation}
\langle V_{\alpha_1}(z_1)...V_{\alpha_n}(z_n)\rangle_{\rm cyl} = \prod_{i<j} \big(2\sin \frac{z_i-z_j}{2}\big)^{\alpha_i\alpha_j}.
\label{fact}
\end{equation}
The factorisation of this correlation simplifies considerably the calculation of $f_n(\alpha,x)$ for the excitation induced by $\Upsilon =e^{i\beta\phi}$. 
Indeed, plugging \eqref{fact} into \eqref{main} and removing the common terms in numerator and denominator, we easily get 
\begin{multline}
f_n(\alpha,x) = \prod_{k=0}^{n-1} \langle V_{\frac{\alpha}{2\pi}}(i\infty)  V_\beta(w_k^{-})  \rangle_{\rm cyl} \prod_{k=0}^{n-1}  \langle V_{-\frac{\alpha}{2\pi}}(-i\infty)    V_\beta(w_k^{-})  \rangle_{\rm cyl} \times \\
 \times  \prod_{k=0}^{n-1}\langle V_{\frac{\alpha}{2\pi}}(i\infty)    V_{-\beta}(w_k^{+})  \rangle_{\rm cyl} \prod_{k=0}^{n-1} \langle V_{-\frac{\alpha}{2\pi}}(-i\infty)    V_{-\beta}(w_k^{+})  \rangle_{\rm cyl}.
\label{Vert}
\end{multline}
%
We regularise our calculation, making the insertion of $V_{\pm\frac{\alpha}{2\pi}}$ at $w=\pm i \Lambda$ and taking the limit $\Lambda \rightarrow \infty$ only at the end. The first two factors of \eqref{Vert}, related to the points $\{w_k^-\}$ at past infinite, give
\begin{multline}
\prod_{k=0}^{n-1} \left( 
\frac{\sin \frac{1}{2} \left( \frac{\pi(1+x)+2\pi k}{n} -i\Lambda \right) }{\sin \frac{1}{2} \left( \frac{\pi(1+x)+2\pi k}{n} +i\Lambda \right)}
\right)^{\frac{\alpha \beta }{2\pi}} 
 \stackrel{\Lambda \to \infty}{\simeq}  \left( e^{i\pi(1+x)}\prod_{k=0}^{n-1}e^{\frac{2\pi i k}{n}} \right)^{\frac{\alpha \beta}{2\pi}} =
  \left( e^{i\pi(1+x)}(-1)^n \right)^{\frac{\alpha \beta }{2\pi}}.
  \label{factor1}
\end{multline}
The other two factors in \eqref{Vert} provide the same result with the replacements $x\rightarrow -x$ and $\beta \rightarrow -\beta$ in Eq. \eqref{factor1}. 
Eventually, multiplying the two, we have the very simple final result
\begin{equation}
f_n(\alpha,x) = e^{i\alpha \beta x}.
\label{Ratio_ver}
\end{equation}
It immediately follows that, when the excited state is induced by the (holomorphic) vertex operator, the only effect is a shift of the mean charge (the 
cumulant generating function is $g_n(\alpha,x)=i \alpha\beta x$).  
In fact, while the average $\bar q$ is not predictable by CFT, its shift from the ground state to a vertex state is universal.
The fluctuations (and all the other cumulants), instead, are the same as those of the ground-state.
The resulting probability function $p(q)$ is shown in Fig.~\ref{prob1}, together with the numerical results for a free fermion model that will be described in the
following (cf. Sec. \ref{num_results}).

For the symmetry resolved R\'enyi entropies, Eq. \eqref{Ratio_ver} implies that $S_n^{V_{\beta}} (q-\bar q_\beta) = S_n^{GS} (q -  \bar q_{GS} )$ 
(where $\bar q_\beta=\bar q_{GS}+ \beta x$ and $\bar q_{GS}$ are the mean values of $Q_A$ in the vertex and ground state respectively). 
In particular, since  equipartition holds for the leading CFT terms, it remains  valid for these excited states.

\subsection{Derivative operator}

In this subsection we consider the other primary operator of the compact boson, namely $\Upsilon= i \partial \phi$.
For the function $f_n (\alpha,x)$ generated by the derivative, we need a general expression for the correlation function
\begin{equation} \label{fUder}
\langle V_{\alpha}(\zeta_1)V_{\beta}(\zeta_2)(i\partial \phi)(z_1)...(i\partial \phi)(z_{2n})\rangle_{\rm cyl},
\end{equation}
for $\beta=-\alpha$. A very useful and standard trick to calculate this kind of correlations is to exploit the identity 
\begin{equation}
(i\partial \phi)(z) = \frac{1}{\epsilon} \partial_z V_{\epsilon}(z)\Big|_{\epsilon=0},
\end{equation}
which allows us to rewrite the desired correlation \eqref{fUder} in terms of derivatives of correlation function of vertex operators in Eq. \eqref{fact} as
\begin{equation}\label{fUder2}
\frac{1}{\epsilon_1...\epsilon_{2n}}\partial_{z_1}...\partial_{z_{2n}}\langle V_{\alpha}(\zeta_1)V_{\beta}(\zeta_2)V_{\epsilon_1}(z_1)...V_{\epsilon_{2n}}(z_{2n})\rangle\Big|_{\epsilon_j=0}.
\end{equation}
For any given $n \in \mathbb{N}$, Eq.~\eqref{fUder2} can be explicitly evaluated. However, in view of the analytic continuation to  non integer $n$, 
we are looking for the general expression as a function of $n$ which is not easily read off from Eq.~\eqref{fUder2}.
We temporarily fix $K=1$ in order to have more compact formulas during the course of the calculation.
In the final result it is enough to replace $\alpha$ with $\alpha \sqrt{K}$ to get the result for generic $K$.
In order to understand the general structure of this correlator, it is instructive to look first at the simplest case $n=1$ which is deduced by the four-point function of 
the vertex operator; after taking the derivatives and the limits $\epsilon_i\to0$,  it reads
\begin{multline}
\langle V_{\alpha}(\zeta_1)V_{\beta}(\zeta_2)(i\partial \phi)(z_1)(i\partial \phi)(z_2)\rangle 
 =\langle V_{\alpha}(\zeta_1)V_{\beta}(\zeta_2)\rangle \times  \\
\times \bigg[  \langle(i\partial \phi)(z_1)(i\partial \phi)(z_2)\rangle  
 + \frac{\alpha^2}{4\tan(\frac{\zeta_1-z_1}{2})\tan(\frac{\zeta_1-z_2}{2})} + \frac{\beta^2}{4\tan(\frac{\zeta_2-z_1}{2})\tan(\frac{\zeta_2-z_2}{2})}+\\
 + \frac{\alpha\beta}{4\tan(\frac{\zeta_2-z_1}{2})\tan(\frac{\zeta_1-z_2}{2})} + \frac{\alpha\beta}{4\tan(\frac{\zeta_1-z_1}{2})\tan(\frac{\zeta_2-z_2}{2})}   \bigg] .
\end{multline}
We now make the following  observations
\begin{itemize}
\item the contribution involving $\zeta_1$ and $\zeta_2$ factorises;
\item the contributions which involve $\zeta_1$ and $z_i$, i.e. $\frac{\alpha}{2\tan(\frac{\zeta_1-z_i}{2})}$, come from $\frac{1}{\epsilon_i}\partial_{z_i}\langle V_{\alpha}(\zeta_1)V_{\epsilon_i}(z_i)\rangle$ (and similarly for $\zeta_2$ with $\alpha\rightarrow \beta$);
\item the term involving $z_i$ and $z_j$ ($z_i\neq z_j$), i.e. $\langle(i\partial \phi)(z_i)(i\partial \phi)(z_j)\rangle=\frac{1}{4\sin(\frac{z_i-z_j}{2})^2}$, 
comes from the derivative $\frac{1}{\epsilon_i \epsilon_j}\partial_{z_i}\partial_{z_j}\langle V_{\epsilon_i}(z_i)V_{\epsilon_j}(z_j)\rangle $;
\item every $z_i$ is connected to another $z_j$ or to $\zeta_j$.
\end{itemize}
It is clear that these observations done at $n=1$ are actually true for any $n$ and follow directly from the factorisation of the vertex correlation function \eqref{fact} 
and the product rule of the differentiation. 
Thus, for general $n$, summing up all the ways that all the points can be connected with the rules above, we obtain the desired correlation function. 
Moreover, we are interested in the case $\zeta_1=i\infty$, $\zeta_2 = -i\infty$ and $\beta = -\alpha$, when the contributions connecting $\zeta_i$ and $z_i$ simplify as follows
\begin{equation} \label{simplify}
\frac{\alpha}{2\tan\frac{i\infty-z}{2}} = 
-\frac{\alpha}{2\tan\frac{-i\infty-z}{2}} = -i\frac{\alpha}{2}.
\end{equation}
Putting all these combinatorial pieces together, it is easy to realise that the desired correlation function can be written as 
\begin{equation} \label{derPM}
\langle V_{\alpha}(i\infty)V_{-\alpha}(-i\infty)(i\partial \phi)(z_1)...(i\partial \phi)(z_{2n})\rangle = \langle V_{\alpha}(i\infty)V_{\beta}(-i\infty)\rangle P_M(i\alpha),
\end{equation}
where $P_M(\lambda)$ is the  characteristic polynomial of the matrix $M$ with elements
\begin{equation} \label{matrixM}
M_{ij} \equiv \begin{cases} \frac{1}{2\sin(\frac{z_i-z_j}{2})} &\quad i\neq j, \\ 0 &\quad i=j.\end{cases}
\end{equation}
From a rigorous point of view Eq. \eqref{derPM} may also be proven by induction, but this is not very instructive. 
Notice that for $\alpha=0$, it reduces to the result for the entanglement entropies in the derivative state obtained in Refs. \cite{excSierra,excSierra2}.
%
Although the expression for $P_M(\lambda)$ is direct and implemented simply for any finite integer $n$, 
it is desirable to write down a more explicit  form that eventually can be analytically continued. 
Indeed, such explicit expression can be obtained by generalising the calculation for the same characteristic polynomial at $\lambda=0$, i.e. $P_M (0)= \det(M)$
presented in Ref.~\cite{det}.  The calculation is cumbersome but straightforward; hence the details of the derivation are reported in Appendix \ref{appendixA}. 

Combining the results in the appendix for $P_M(i\alpha)$ with the other correlation functions entering in $f_n(\alpha,x)$, cf. Eq. \eqref{main}, 
it reduces to the following  polynomial of degree $2n$ for any integer $n$:
\begin{equation}
f_n(\alpha,x) = \prod_{p=1}^n \left(   1-\left(  \frac{\alpha}{\pi} \right)^2 \frac{1}{(\frac{n}{\sin(\pi x)} -n-1+2p)^2}        \right).
\label{pol}
\end{equation}
This expression is easily analytical continued to arbitrary non integer values of $n$ as
\begin{multline}
f_n(\alpha,x) = \left(\frac{\Gamma(1+n + \frac{1}{2}(\frac{n}{\sin(\pi x)} -n-1) +\frac{\alpha}{2\pi})}{\Gamma(1+ \frac{1}{2}(\frac{n}{\sin(\pi x)} -n-1)+\frac{\alpha}{2\pi})} \right)  \times \\ 
\times \left(\frac{\Gamma(1+n + \frac{1}{2}(\frac{n}{\sin(\pi x)} -n-1) -\frac{\alpha}{2\pi})}{\Gamma(1+ \frac{1}{2}(\frac{n}{\sin(\pi x)} -n-1)-\frac{\alpha}{2\pi})} \right) 
  \left(\frac{\Gamma(1+ \frac{1}{2}(\frac{n}{\sin(\pi x)} -n-1))}{\Gamma(1+n + \frac{1}{2}(\frac{n}{\sin(\pi x)} -n-1))} \right)^2,
\label{res2}
\end{multline}
where we just used repeatedly the analytic  continuation of the factorial $\Gamma(n+1)=n!$.

\begin{figure}[t]
\centering
\includegraphics[width=0.46\textwidth]{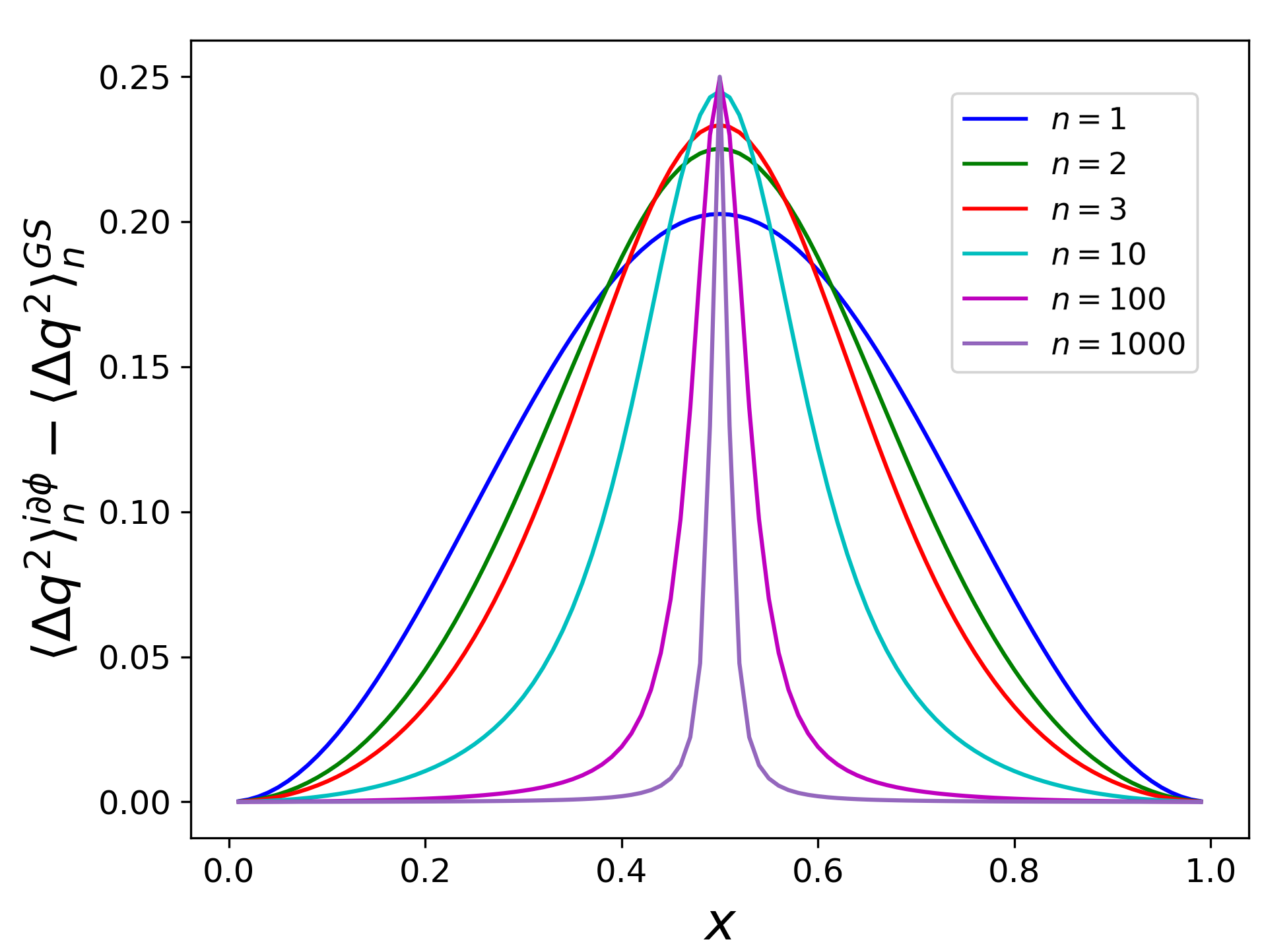}
\includegraphics[width=0.484\textwidth]{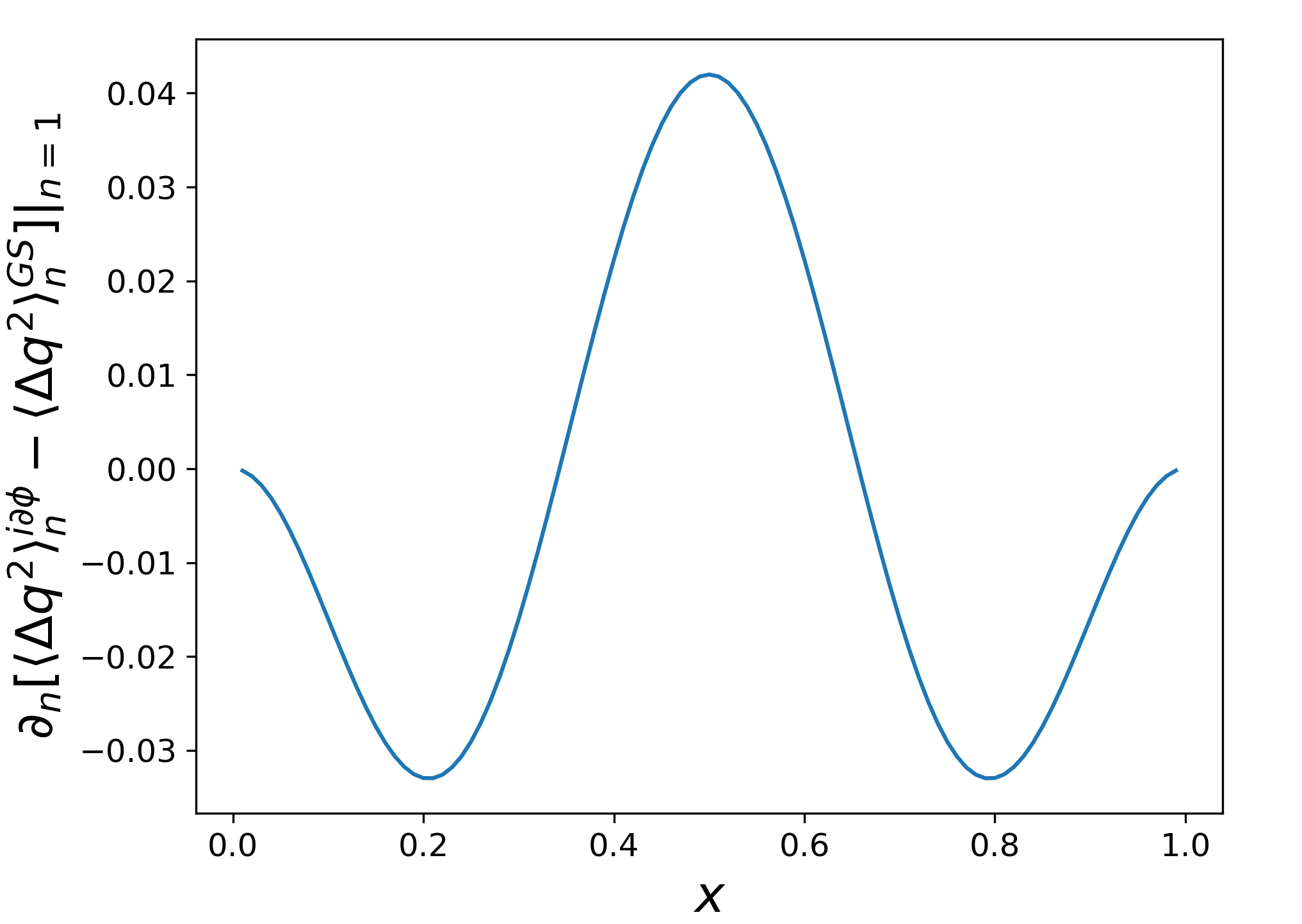}
\caption{Left: Excess of variance \eqref{VarDer} for the derivative state compared to the ground state as function of $x=\ell/L$. 
The curves correspond to $n=1,2,3,10,100,1000$. Notice as they quickly shrinks as $n$ increases. 
Right: Derivative wrt to $n$ at $n=1$ of the excess of variance as a function of $x$: importantly this function is not monotonic. 
 }
	\label{Vn}
\end{figure}

Eq. \eqref{res2} provides all the connected moments of the generalised probability $p_n^{i \partial \phi} (q)$ as in Eq. \eqref{cumuk};
first the shift of the mean value vanishes $\langle q \rangle^{i\partial \phi}_n-\langle q \rangle^{GS}_n=0$; 
then, the excess of variance is 
\begin{multline}
\langle \Delta q^2\rangle^{i\partial \phi}_n -\langle \Delta q^2\rangle^{GS}_n   = - \frac{d^2}{d\alpha^2} g_n(\alpha,x)\Big|_{\alpha=0}=
\frac1{\pi^2} \sum_{p=1}^n \Big( \frac{n}{\sin \pi x} -n-1+2p \Big)^{-2}=\\
 =  \frac{1}{2\pi^2}\left( \psi\left( \frac{1}{2} \left( \frac{n}{\sin(\pi x)} -n-1\right)+1 \right) -\psi\left( \frac{1}{2}\left( \frac{n}{\sin(\pi x)} -n-1\right)+1+n\right)\right),
\label{VarDer}
\end{multline}
where the first line is valid for integer $n$ and the second one is the analytic continuation in terms of the digamma function
$\psi(z)$ (the logarithmic derivative of the $\Gamma$ function). 
Note that, as already stressed for a general state (cf. Eq. \eqref{cumuk}),  the variance excess is universal for any $n$. 
More generally all the differences of cumulants between the excited states and ground state are universal (cf. Eq. \eqref{cumuk}).
When $n=1$, we have the very compact result
\begin{equation}
\sigma_1^2 \equiv\langle \Delta q^2\rangle^{i\partial \phi}_1 -\langle \Delta q^2\rangle^{GS}_1 = \frac{2}{\pi^2}\sin(\pi x)^2,
\label{sigma1}
\end{equation}
which has a direct physical meaning as fluctuations of the charge (this is valid only for $K=1$, for different values Eq. \eqref{sigma1} gets multiplied by $K$). 
Eq. \eqref{sigma1} implies that the state corresponding to $\Upsilon = i\partial \phi$ has fluctuations that are not read off from the ground-state 
 as it is instead the case for the vertex operator. 
The excess of variance as predicted by Eq. \eqref{VarDer} is reported in Figure \ref{Vn} (left panel) as a function of $x$ for various $n$.
The behaviour as a function of  $n$ is rather peculiar since the various curves cross, signalling a non uniform and monotonic behaviour in $n$ and $x$.
Notice that the curves shrink quickly as $n$ increases and, in fact, the limit for $n\to\infty$ is a discontinuous function equal to $0$ for all $x$, except for 
$x=1/2$ when the limit is $1/4$. This discontinuous function is expected to lead to very strong  finite size effects .
In the right panel we report the derivative of the excess of variance wrt to $n$ at $n=1$. 
This function, as we shall see, enters in the symmetry resolved von Neumann entropy. 
It is non-monotonic in the interval for $x\in [0,1/2]$ and it changes sign, a rather unusual shape which leads to non uniform finite size corrections.

\subsubsection{From charged moments to symmetry resolution}

From the knowledge of the universal function $f_n(\alpha,x)$, we straightforwardly get the generalised moment distribution function   
$p_n^{i\partial \phi}(\alpha,x) = p_n^{GS}(\alpha,x)f_n(\alpha,x)$, cf. Eq. \eqref{fnpns}. 
However, the computation of the symmetry resolved entanglement entropies $S_n(q)$ requires the knowledge of its Fourier transform, $p_n^{i\partial \phi}(q)$. 
Again, for conciseness of the formulas, we set $K=1$ hereafter. 
For different $K$, the rescaling  $\alpha\to \sqrt{K} \alpha$, leads to $p_n(\Delta q) \to  p_n(\Delta q /\sqrt{K})/\sqrt{K}$. 

\begin{figure}[t]
\centering
\includegraphics[width=0.5\textwidth]{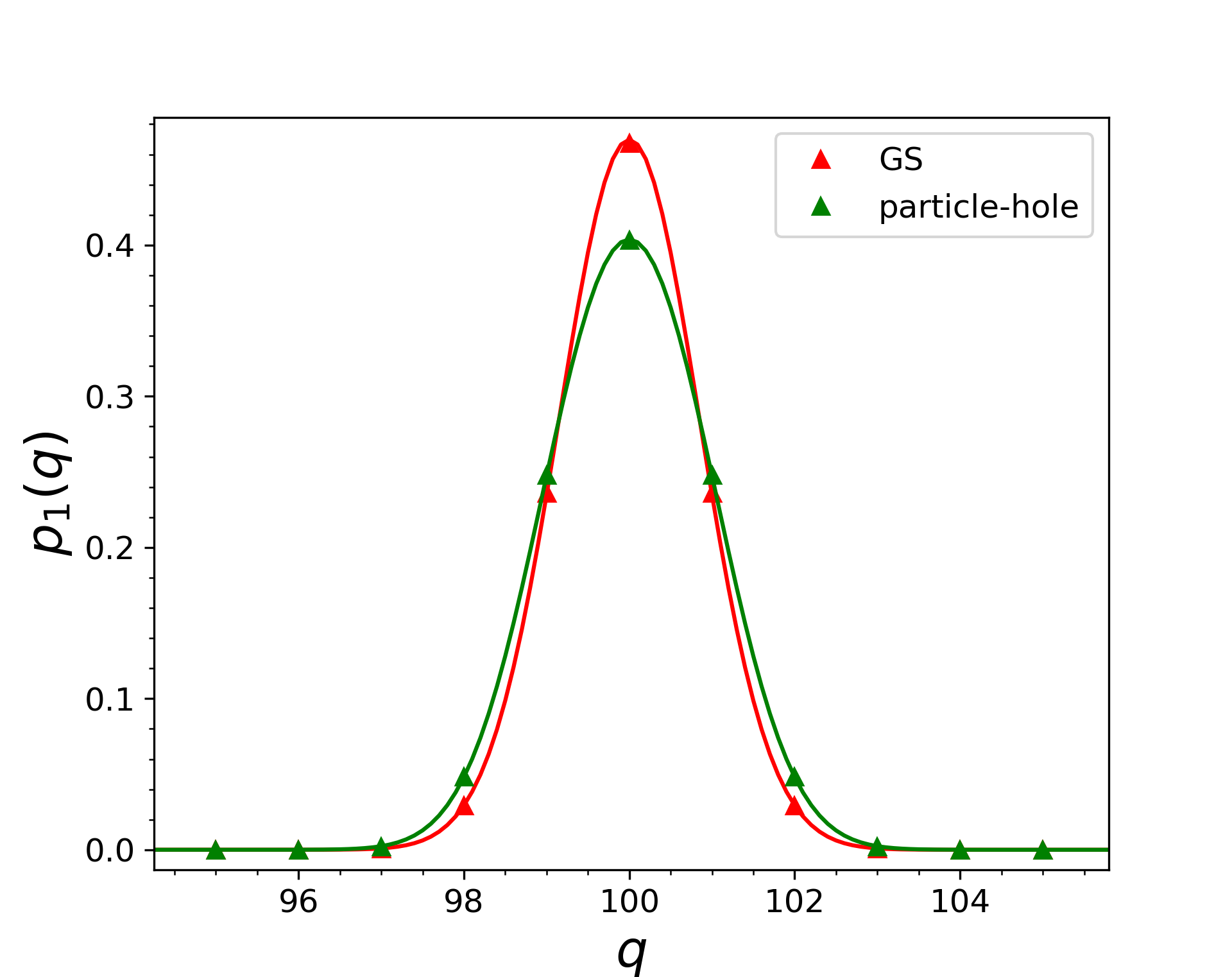}
\caption{
Probability  distribution of the number of particles in the ground state and in the excited state induced by the derivative operator $i \partial \phi$. 
The CFT prediction \eqref{nongauss} is shown as a full line. 
The points are numerical data  for the half-filled XX chain, cf. Eq. \eqref{Hamiltonian}, with length $L=400$ and subsystem of size $\ell=L/2$.
The excitation is a particle-hole created over the Fermi sea in the lattice model.
The non-universal constants are fixed as in Figure \ref{prob1}.
While for the ground-state the probability distribution is Gaussian, for the excited state this is no longer the case.}
	\label{prob2}
\end{figure}

It is instructive to explore first what happens for $n=1$. 
From Eq. \eqref{pol} we read the generating function
\begin{equation}
p_1^{i \partial \phi}(\alpha,x) = e^{-\frac{\alpha^2\sigma_0^2}{2}}\left(1-\frac{1}{2}(\sigma_1\alpha)^2\right),
\label{paD}
\end{equation}
where $\sigma_0^2 = \langle \Delta q^2\rangle_1^{GS}$ and $\sigma_1$ is in Eq. \eqref{sigma1}.
This generating function is even, so all odd cumulants are zero. 
Instead,  the excess of even cumulants $\kappa_k^{i\partial \phi}$ with $k\geq 4$ is
\begin{equation}
\kappa_{2k}^{i\partial \phi}-\kappa_{2k}^{GS}= (-1)^{k-1} \frac{(2k)!}{2^k k} \sigma_1^{2k},\qquad k\geq 2,
\label{cumex}
\end{equation}
which are non zero, universal, and of order one. 
We recall that in the ground state within CFT only  the variance is non zero because the probability $p_1^{GS}(q)$ is Gaussian.
However, any microscopic model has $O(1)$ non-universal cumulants (see e.g. \cite{Laf,Min}); 
the meaning of Eq. \eqref{cumex} is that $\kappa_{2k}^{i\partial \phi}-\kappa_{2k}^{GS}$ is 
universal although it is a difference of two $O(1)$ term. 
{\it The non-Gaussian nature of $p_1^{i\partial \phi}(q)$ and the universality of the difference of higher cumulants}
 is an important observation, that, to the best of our knowledge, has not been done in the past. 
Unfortunately, since the variance of the distribution is dominated by the ground-state value $\sigma_0^2$ which is proportional to $\log \ell$,
all these nice $O(1)$ universal non-gaussian effects are subleading corrections to the ground-state gaussian behaviour
(this is highlighted by the behaviour of the kurtosis scaling  as $\propto  (\log \ell)^{-2}$). 
Taking the Fourier transform of Eq. \eqref{paD}, we get the probability
\begin{equation}
p_1^{i \partial \phi}(q) = \frac{1}{\sqrt{2\pi \sigma_0^2}}e^{-\frac{\Delta q^2}{2\sigma_0^2}}\left( 1-\frac{1}{2}\left( \frac{\sigma_1}{\sigma_0} \right)^2 +\frac{\Delta q^2\sigma_1^2}{2\sigma_0^4} \right),
\label{nongauss}
\end{equation}
which clearly is not Gaussian. 
This is shown graphically in Fig.~\ref{prob2} (together with the data for a real system, discussed in the next subsection, to report realistic values for the non-universal constants). 
Anyhow, the leading deviation from Gaussianity is proportional to $(\sigma_1^2/\sigma_0^2)\propto (\log \ell)^{-2}$: 
although this is only a subleading correction to the scaling for large $\ell$, it decays extremely slowly with $\ell$ and its effects are rather strong and visible also in Fig.~\ref{prob2}.

Let us now move to a generic $n$. When $n \in \mathbb{N}$, we have shown that $f_n(\alpha,x)$ is an even polynomial in $\alpha$ (cf. Eq. \eqref{pol}).
We then just need the Fourier transform of $e^{\frac{1}{2}\alpha^2 \sigma_0^2}\alpha^{2k}$, which is 
\begin{equation}
\int_{-\infty}^{\infty} \frac{d\alpha}{2\pi} e^{-i\alpha q}e^{-\frac{\alpha^2 \sigma_0^2}{2}}\alpha^{2k} = 
\frac{e^{-\frac{q^2}{2\sigma_0^2}}}{\sqrt{2\pi \sigma_0^2}} (-1)^{k} H_{2k} \left( \frac{q}{\sqrt{2}\sigma_0} \right) \frac{1}{(\sqrt{2}\sigma_0)^{2k}},
\label{FT}
\end{equation}
where $H_{2k}(x)$ is the $2k$-th Hermite polynomial. 
Thus, as long as $n$ is integer, $p_n^{i\partial \phi}(q)$ is the sum of a finite number of terms of the form of the Fourier transforms in Eq. \eqref{FT} 
and all the universal even higher cumulants can be calculated although their full analytic expression is unwieldy for $n>1$.
For fixed non-integer $n$,  we can numerically perform the Fourier transform of $p_n(\alpha)$ to extract the probability $p_n(q)$ and from this the symmetry resolved 
entanglement of our interest, but we do not get close explicit expressions for it. 
The main drawback of this was of proceeding is that we do not have an analytic formula to perform the analytic continuation for the von Neumann entropy.

A possible approximation to handle analytically the problem is to keep only the first order $\alpha^2$ in the expansion of $f_n(\alpha)$, i.e. to approximate the generating function as
\begin{equation}
p_n^{i\partial \phi}(\alpha,x) \simeq e^{-\frac{\alpha^2}{2}\langle \Delta q^2\rangle_n^{GS}}\left(1-\frac{\alpha^2}{2}(\langle \Delta q^2\rangle_n^{i\partial \phi}-\langle \Delta q^2\rangle_n^{GS} )\right).
\label{q_app}
\end{equation}
This approximation has the advantage that is well defined for any $n$, even non integer. 
It is also motivated by the fact that the neglected contributions $\propto \alpha^{2k}$, after Fourier transform, in the symmetry resolved entropies 
provide terms which are suppressed as higher powers of $(\log \ell)^{-1}$. 
Unfortunately, such log-corrections are for generic $n$ too slow to be ignored. 
However, close to $n=1$, this quadratic approximation works very well since exactly at $n=1$ it becomes exact, see Eq. \eqref{paD}.  
Within this approximation, we obtain an analytic result for the symmetry resolved entanglement entropies.
The resulting expression is rather cumbersome. 
Thus, to lighten the notation, we define $d_n \equiv \langle \Delta q^2\rangle_n^{i\partial \phi}-\langle \Delta q^2\rangle_n^{GS}$. 
In the physical regime $\Delta q$ of order $1$, we then have
\begin{equation}
S_n^{i\partial \phi}(q) = \Delta S+S_n^{GS}(q) +\frac{1}{n-1}\frac{\Delta q^2}{2(\frac{K}{n\pi^2} \log \ell)^2}\left(\frac{d_1}{n}-d_n\right)+
o((\log \ell)^{-2}).
\label{Snq}
\end{equation}
Here $\Delta S$ is $q$ independent; its precise form is not very illuminating but we report it anyway:
\begin{equation}
\Delta S=S_n^{i\partial \phi}- S_n^{GS}+\frac{1}{1-n}\frac{1}{2 \frac{K}{n\pi^2} \log \ell}\left(d_1-d_n\right) +
\frac{1}{1-n}\frac{1}{( \frac{K}{n\pi^2} \log \ell)^2} \left(\frac{d_nb_n}2-\frac{d_1b_1}{2n} -\frac{d_n^2}{8} +\frac{d_1^2}{8n} \right).
\end{equation}
We stress once again that the approximation \eqref{Snq} is not very effective for $n > 1$ at the value of $\ell$ that are usually accessed by numerical calculations, 
but works very well at $n=1$, as we shall see. 
A very important aspect of Eq. \eqref{Snq} is the presence of a {\it universal} term that breaks equipartition of entanglement at order $(\log \ell)^{-2}$. 
Hence, the term breaking equipartition is of the same order as the non-universal cutoff term in the variance, cf. Eq. \eqref{Dom}. 
However,  the latter may be subtracted considering differences with the ground state values.

\subsection{Numerical tests for free fermionsl} \label{num_results}

We now provide numerical tests of the universal CFT results in the previous subsections, using free-fermion techniques \cite{peschel2001,peschel2003, pe-09}. 
We consider the tight-binding model, i.e., a 1D chain of free fermions described by the following hamiltonian
\begin{equation}
H = - \sum_j \left[c_j^\dagger c_{j+1} + c_{j+1}^\dagger c_j -2h \Big( c_j^\dagger c_j -\frac{1}{2} \Big) \right],
\label{Hamiltonian}
\end{equation}
with $c_j^\dagger$, $c_j$ a set of lattice fermionic ladder operators, satisfying the anticommutation relations $\{ c_i, c_j^{\dagger}\} = \delta_{ij}$ and $h$ is the chemical potential.
It is well known that by a Jordan-Wigner transformation, this model is mapped to the XX spin chain \cite{sach-book} and that (being the Jordan-Wigner local within a block) 
the fermion entanglement is the same as the spin one \cite{atc-10,ip-10}. For this reason, we will also refer to the Hamiltonian \eqref{Hamiltonian} as the XX spin chain.  
The ground state of the Hamiltonian \eqref{Hamiltonian} is a Fermi sea with Fermi momentum $k_F = \arccos{|h|}$. 
The $U(1)$ symmetry is related to the conservation of the number of fermions $N=\sum_j c_j^\dagger c_j$. 
We are interested in the spatial bipartition of the system where $A$ is given by $\ell$ contiguous lattice sites. 
The RDM is Gaussian and it can be written as \cite{pe-09}
\begin{equation}
\rho_A  = \det C_A \exp (\sum_{i,j} \log [(C_A^{-1}-1)]_{ij}c_i^\dagger c_j),
\label{eham}
\end{equation}
where the $\ell \times \ell$ matrix $C_A \equiv \langle c_i^\dagger c_j\rangle$ (with $i,j\in A$) is the \textit{correlation matrix} restricted to $A$. 

For an infinite chain ($L=\infty$), in the ground state, $C_A$ has the following elements
\begin{equation}
(C_A^{GS})_{ij} = \frac{\sin k_F(i-j)}{\pi(i-j)}.
\end{equation}
However, in our case, we are interested in finite $L$, when the excited states have a finite excess of entropy (which instead vanishes in the thermodynamic limit). 
In this case, it holds
\begin{equation} \label{CAgsL}
(C_A^{GS})_{ij} = \frac{1}{L}\frac{\sin k_F(i-j)}{\sin\frac{\pi(i-j)}{L}}.
\end{equation}
Using \eqref{eham} together with \eqref{CAgsL}, a system with $2^\ell$ degrees of freedom can be studied through the numerical diagonalization of a $\ell \times \ell$ matrix, a low demanding numerical task, especially when compared to the exact diagonalisation of the entire Hamiltonian. 

\begin{figure}[t]
\centering
\includegraphics[width=0.6\textwidth]{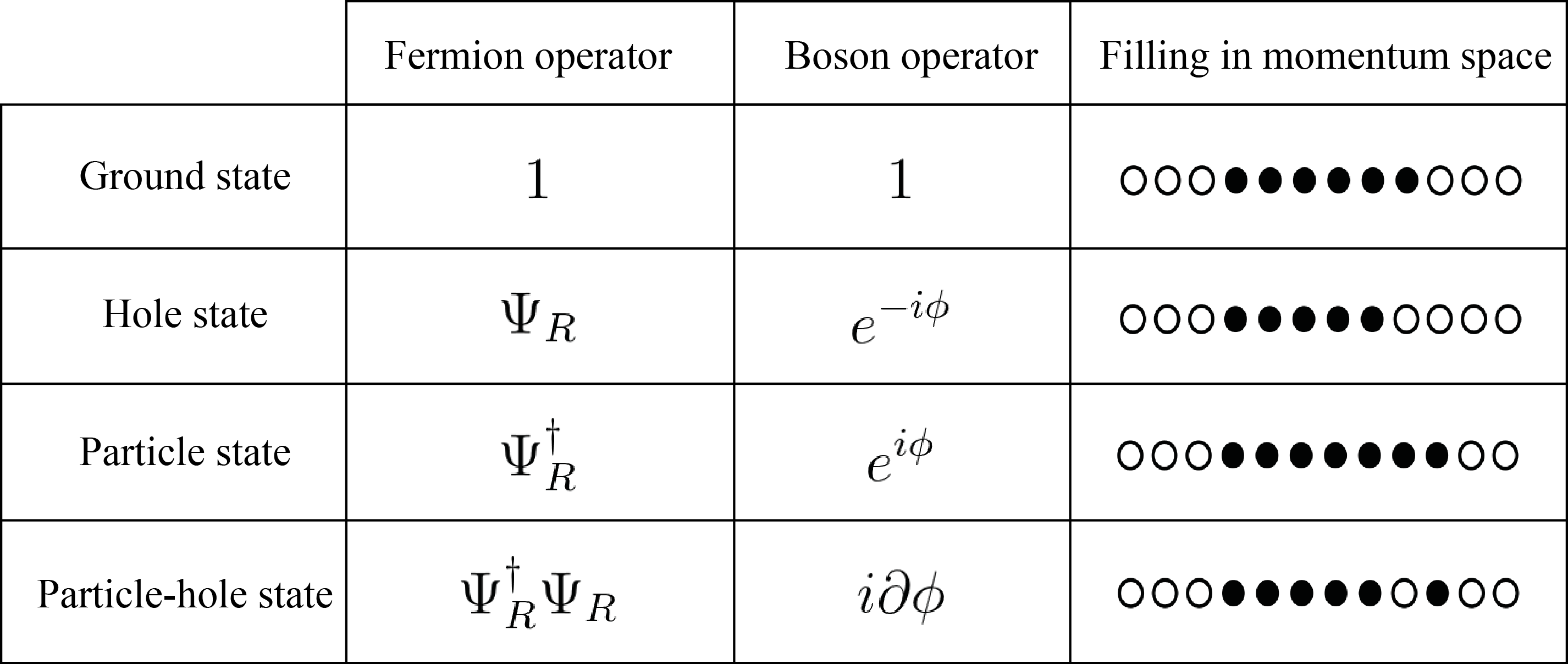}
\caption{Bosonisation dictionary for the low-energy excitations of the free-fermion chain with Hamiltonian \eqref{Hamiltonian}. 
In this notation $\Psi_R$ ($\Psi_R^\dag$) is the annihilation (creation) operator of the continuum theory at the right Fermi momentum (on the lattice 
it corresponds to $c_{k_F-\pi/L}$ ($c^\dag_{k_F+\pi/L}$)).
}
\label{dic}
\end{figure}

In general, the Wick theorem allows us to apply this method not only to the ground state, but to all excited states (in Fock basis) which are gaussian.
The low lying states are excitations of particles and holes above or below the Fermi sea and correspond to the primary operators of the Luttinger liquid 
via the state-operator correspondence. We briefly recall the bosonization dictionary in Fig.~\ref{dic} (for further details see, e.g., \cite{excSierra}).
The two states we consider are i) the vertex operator $e^{i \phi}$ which corresponds to a particle excitation the (right) Fermi point (cf. Fig.~\ref{dic}) 
with correlation matrix
\begin{equation}
(C_A^{e^{i\phi(z)}})_{jj'} = (C_A^{GS})_{jj'} + \frac{1}{L}e^{-i(k_F + \frac{\pi}{L})(j-j')},
\label{CAV}
\end{equation}
and ii) the derivative operator $i\partial \phi(z)$ which corresponds a (right) particle-hole excitation  (cf. Fig.~\ref{dic})  with correlation matrix
\begin{equation}
(C_A^{i\partial \phi})_{jj'} = (C_A^{GS})_{jj'} + \frac{1}{L}e^{-i(k_F + \frac{\pi}{L})(j-j')}- \frac{1}{L}e^{-i(k_F - \frac{\pi}{L})(j-j')}.
\label{CAD}
\end{equation}
Notice that both Eqs. \eqref{CAV} and \eqref{CAD} in the thermodynamic limit $L\to\infty$ reduce to $C_A^{GS}$ as they should.

\subsubsection{Symmetry resolved moments and their generating function}

In a general state, any local operator within $A$ can be written in terms of $\rho_A$ and hence, thanks to Eq. \eqref{eham}, in terms of $C_A$. 
In particular, the entanglement spectrum is only a function of the spectrum of $C_A$ that we denote as $\{\nu_k\}$. 
For example the total R\'enyi entropies are written as \cite{pe-09}
\begin{equation}
S_n \equiv \frac{1}{1-n} \sum_{k=1}^\ell \log (\nu_k^n+(1-\nu_k)^n  )\,.
\end{equation}
Similarly the charged moments are \cite{Gsela}
\begin{equation}
\text{tr}(\rho_A^n e^{i\alpha Q_A}) = \prod_k(e^{i\alpha}\nu_k^n + (1-\nu_k)^n),
\label{expr}
\end{equation}
where we recall that $Q_A = \sum _{j \in A} c_j^\dagger c_j$.

\begin{figure}[!t]
{\includegraphics[width=0.96\textwidth]{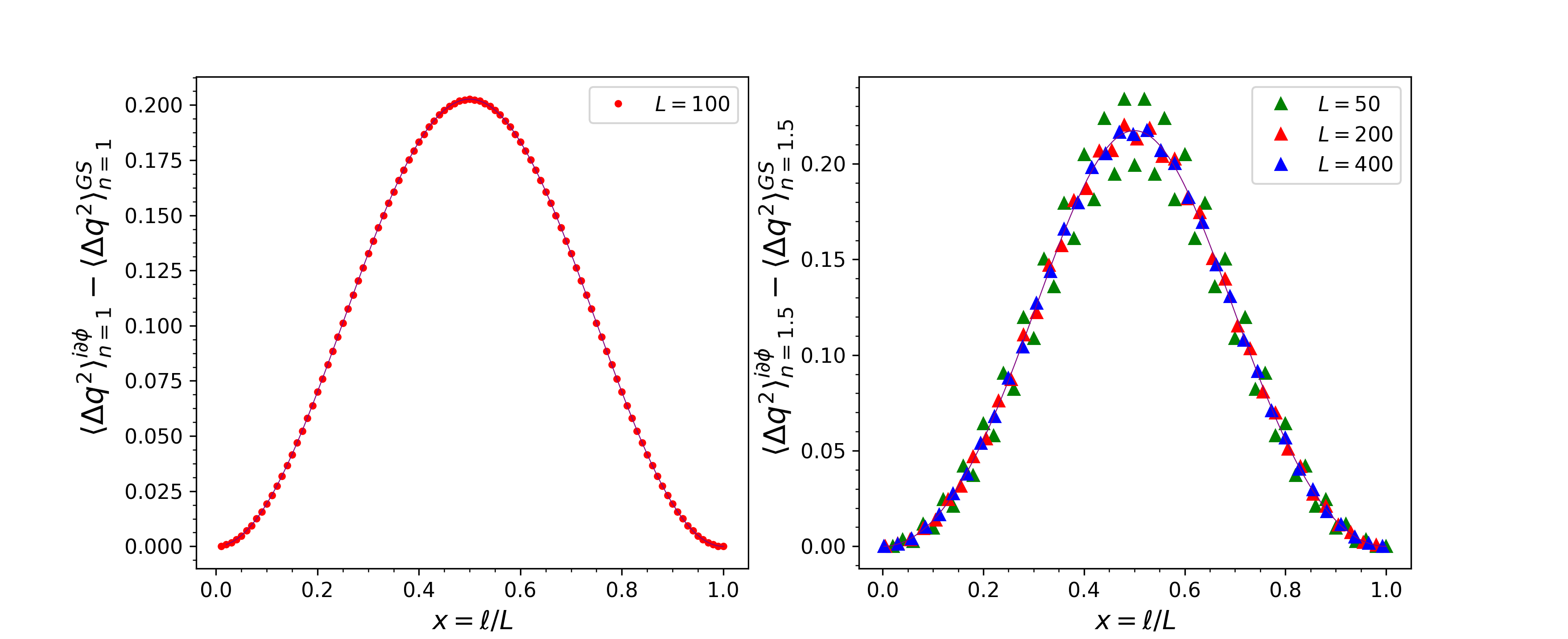}}\qquad
{\includegraphics[width=0.96\textwidth]{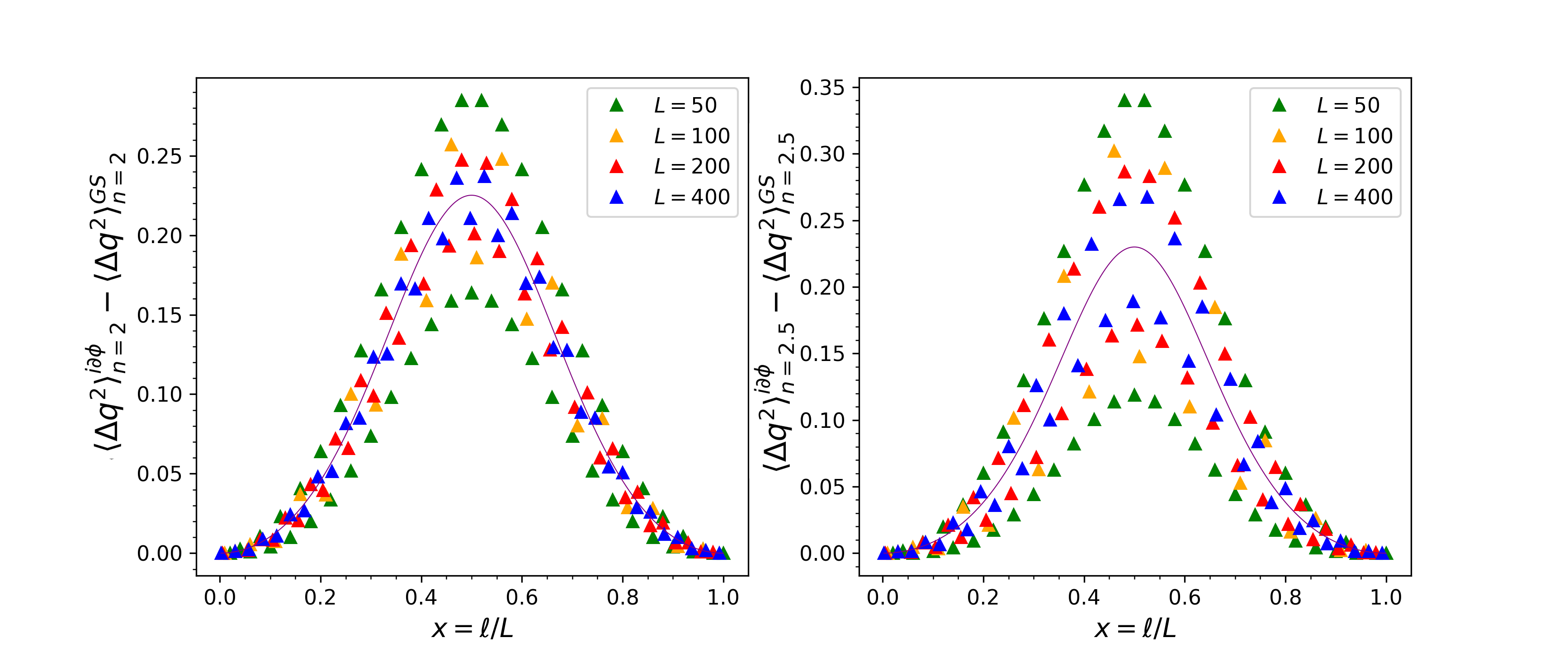}}
\caption{Variance excess of the particle-hole excited state respect to the ground state for different $n=1, 1.5, 2, 2.5 $. 
The numerical data refer to a half-filled XX chain of finite size (different values of $L$ are shown). The continuous lines are the CFT predictions, Eq.~\eqref{VarDer}. 
The excess of variance is maximum when $x=\frac{1}{2}$ ($\ell = \frac{L}{2}$). }
\label{Var_check}
\end{figure}

When calculating  moments and cumulants of the probabilities $p_n(q,x)$, it is not necessary to calculate first the probability using Eq. \eqref{expr}
and, from these, the moments. 
It is more effective to write directly the moments in terms of the eigenvalues $\nu_k$ calculating the derivatives wrt $\alpha$ of the generating function \eqref{expr} 
written as sum over $\nu_k$ (this  is already routinely done for $n=1$, e.g., in Refs. \cite{Laf,Min}).  
As an example,  the variance of $p_n$ is 
\begin{equation}
\langle \Delta q^2 \rangle_n = \sum_k \frac{1}{(\nu_k^{-1} -1)^n+1}-\frac{1}{((\nu_k^{-1} -1)^n+1)^2}.
\end{equation}
Similar formulas for higher moments are straightforwardly written down. 

\begin{figure}[t]
{\includegraphics[width=0.99\textwidth]{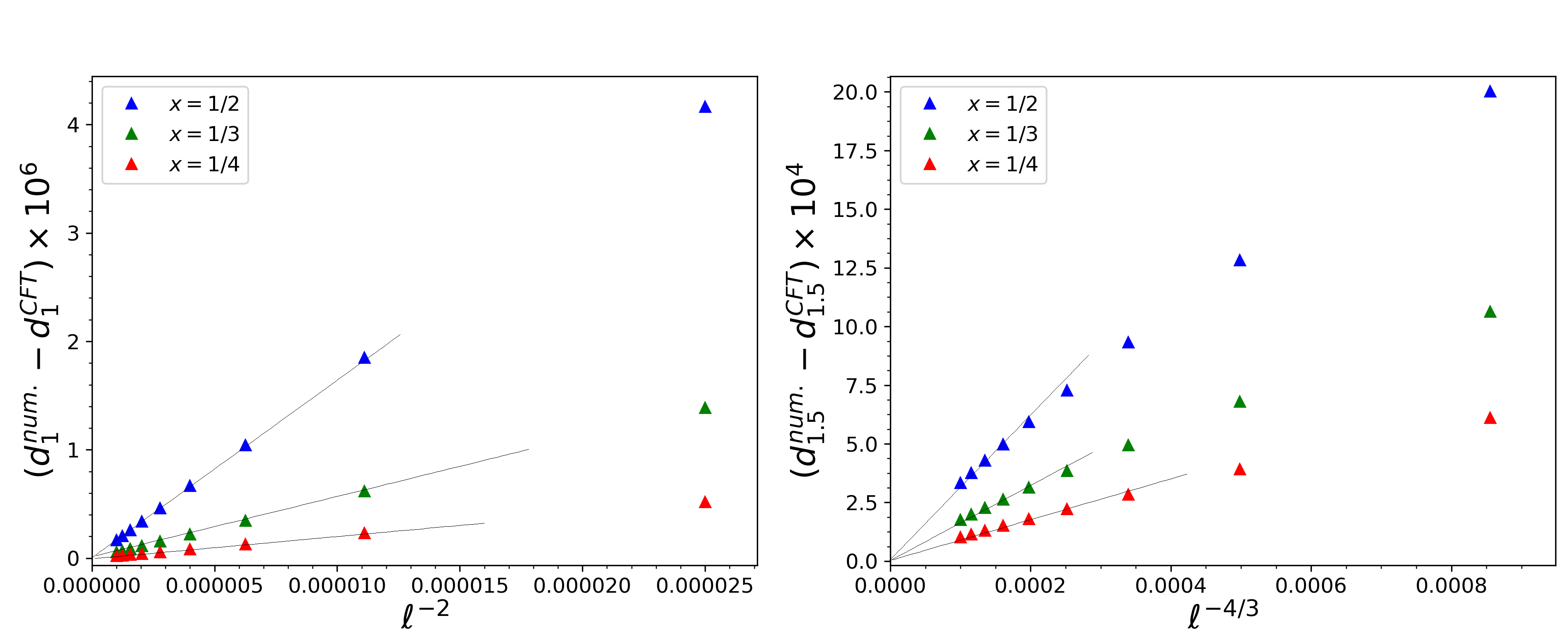}}\
\caption{Corrections to the scaling for the excess of variance $d_n= \langle \Delta q^2\rangle_n^{i\partial \phi}-\langle \Delta q^2\rangle_n^{GS}$ 
for the particle-hole state respect to the ground state. 
We report the numerical data minus the leading CFT prediction.
We plot the data against the expected scaling of the corrections $\ell^{-2/n}$ for $n=1$ (left) and  $1.5$ (right) for three values of the ration $x=\ell/L$.
The straight lines are guide to the eyes with the expected asymptotic behaviour.
}
\label{Fig:corr}
\end{figure}

We start our numerical analysis of the free fermion chain from the variance in the excited state: we focus on the excess of variance between an excited state and the ground state, 
because, as we have shown (cf. Eq. \eqref{cumuk}), it is universal and does not depend on any microscopical detail. 
%
For the vertex operator, the CFT prediction \eqref{Ratio_ver} implies that all moments for any $n$ are the same as in the ground state.
For the free fermion chain, this property trivially follows from the fact that the excited state remains a compact Fermi sea, just  with one particle more 
at the right Fermi point \cite{excSierra2}.
In fact, the operator $e^{i\phi}$ corresponds to a particle excitation in the right sector of the Fermi sea and the meaning of Eq. \eqref{Ratio_ver} is that 
one finds the exceeding particle with a probability ${\ell}/{L}$.
Therefore, we now focus on the excess of variance of the derivative operator. 
The numerical calculated data for $n=1,1.5,2,2.5$ and for several values of $L$ in the range from $L=50$ to $L=400$, are reported as function of $x=\ell/L$ in
Figure~\ref{Var_check}.
The comparison between the CFT prediction~\eqref{VarDer} and numerics is shown in the same picture. 
It is clear that, when increasing the system size, the points get closer to the analytical curve, valid in the thermodynamic limit, in spite of the 
presence of oscillating corrections to the scaling with amplitudes that clearly decrease with system size. 
These oscillations get larger for larger values of the R\'enyi index $n$ and they are absent at $n=1$ when, for $L$ as small as $100$, the numerical 
data are perfectly on top of the CFT prediction.
Such oscillations do not come unexpectedly: they are just the well known {\it unusual corrections} to the scaling.
In the ground state they have been fully characterised both in CFT \cite{sublead,ot-15}, and in microscopic 
models \cite{Fagotti2010cc,parity2,parity,correzioni,mint4,overlap1,num} 
for the total R\'enyi entropies (and are known to be absent, instead, in the case $n=1$, the Von Neumann entropy).
They are present not only in the ground state, but in excited states as well \cite{sublead1,excSierra2}; 
indeed they are related to the structure of conical singularities in CFT \cite{sublead} (i.e. to the Riemann surface ${\cal R}_n$), 
are not affected by possible operator insertions, and so they are independent of the state (the amplitude, however, depends in a complicated 
and yet unknown manner on the state itself). 
In Ref. \cite{Ric, Gold}  through the (generalised) Fisher-Hartwig conjecture, it has been shown that for free fermions such corrections for the charged moments
scale like $L^{-2/n(1-\alpha/\pi)}$ replacing the well known decay $L^{-2/n}$ at $\alpha=0$; 
hence they become larger as $\alpha$ moves away from zero (and the CFT argument of Ref. \cite{sublead}
is easily modified to predict such new decay). 
Anyhow, since the variance is defined at $\alpha=0$, the corrections in Figure~\ref{Var_check} should decay as $L^{-2/n}$.
This scaling is explicitly tested in Figure \ref{Fig:corr}, where we show that the difference between numerical data and CFT prediction for the 
excess of variance  indeed decays as $\ell^{-2/n}$ (at fixed $x$ we can replace $L$ with $\ell$). 

\begin{figure}[!t]
\includegraphics[width=1\textwidth]{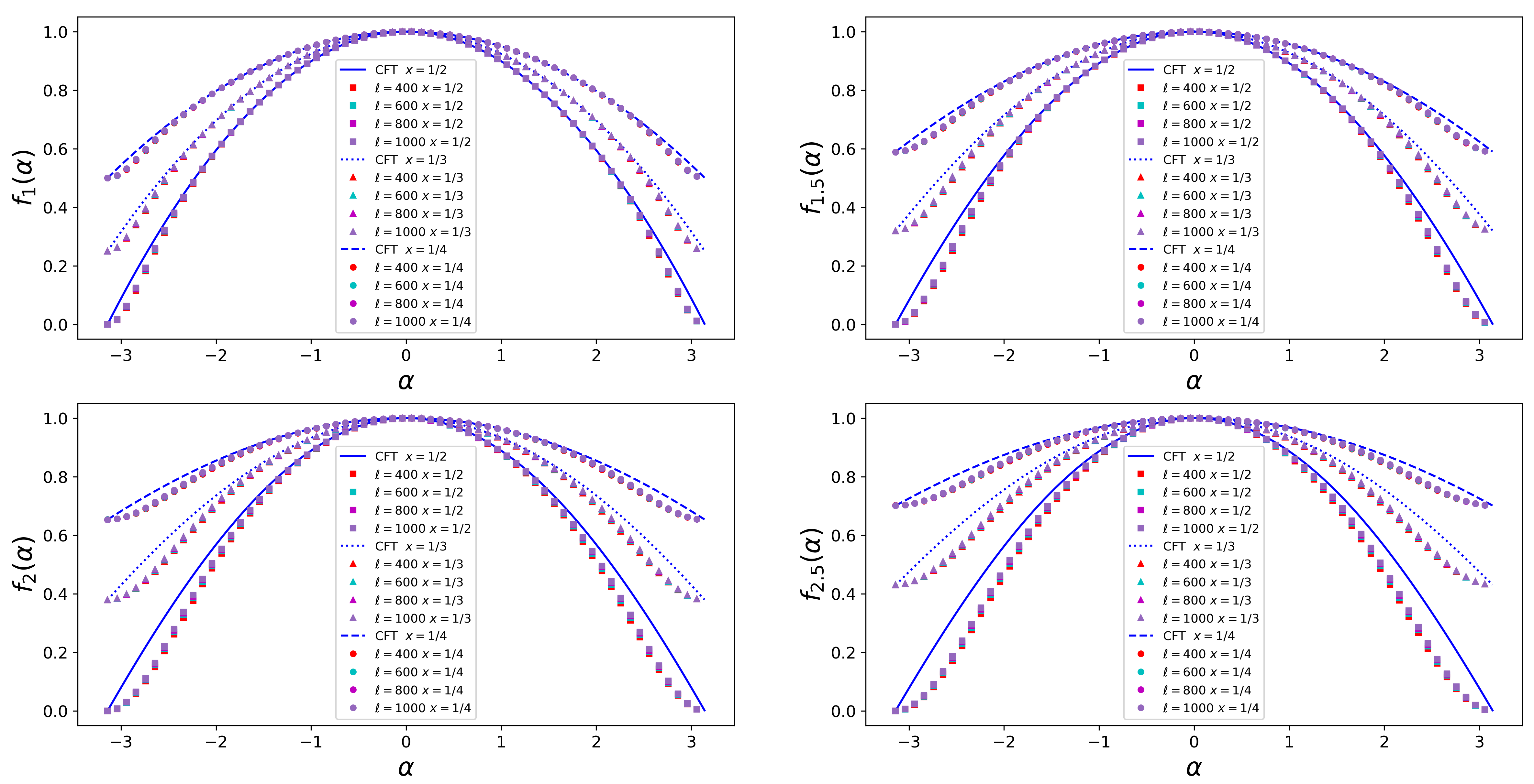}
\caption{Numerical data for $f_n(\alpha,x)$ for the particle-hole excitation in the XX chain (symbols) compared with the CFT prediction for the derivative operator, 
cf. Eqs.~\eqref{pol} and \eqref{res2}. We report results for $n=1,1.5,2,2.5$ and $x=1/4, 1/3, 1/2$. We show the data for several values of $\ell$ up to 1000.
The agreement is very good for small $\alpha$, but it worsens as $\alpha$ gets closer to $\pm\pi$ and as $n$ gets larger.}
\label{fna}
\end{figure}

We now move to the analysis of the charged moments or generalised cumulants generating functions $p_n(\alpha,x)$. 
Again for the vertex operator they are trivially equal to the ground state ones, apart from a phase (see Eq. \eqref{Ratio_ver} and Fig. \ref{prob1}). 
Hence, we focus here on the non-trivial the derivative operator.
The numerical results for the function $f_n (\alpha,x)$ for $n=1,1.5,2,2.5$ are reported in Fig.~\ref{fna} 
(for $n=1$, these are just the data for the generating function of the full counting statistics of the charge which, surprisingly,   
has not yet been considered in the literature). 
The agreement between CFT prediction \eqref{res2} and numerical data is excellent at small $\alpha$, while it gets worse for larger values of $\alpha$ and $n$. 
This is not surprising; as already discussed, in the ground state the corrections to the scaling decay as $L^{-2/n(1-\alpha/\pi)}$ \cite{Ric, Gold} becoming larger 
as $\alpha$ moves away from zero; the same remains true for excited states since the insertion of operators does not alter the structure of the Riemann surface
(as the flux does \cite{Gsela}).
In fact, even in the thermodynamic limit ($\ell, L \rightarrow \infty$, with $x$ kept fixed) one expects convergence only in the region $\alpha \in [-\pi,\pi]$; 
on a lattice with lattice spacing $a$ (which we set to 1) by definition (cf. Eq. \eqref{expr}) $f_n(\alpha,x)$ is periodic with period $2\pi/a$, but this cannot be captured by the CFT
working in the limit $a\to0$; the entire $f_n(\alpha,x)$ for any $\alpha\in \mathbb R$ can be reconstructed  by  periodically continuing it outside the domain 
$[- \pi, \pi]$ (see \cite{Gold} for details). 
Anyhow, this effect does not affect the behaviour of the cumulants of the charge $Q_A$ that are obtained as derivatives with 
respect to $\alpha$ evaluated at $\alpha=0$. 
Finally, we recall that for $n\to\infty$, $f_n(\alpha,x)$ becomes discontinuous inducing large finite size effects for large $n$.

\begin{figure}[!t]
\includegraphics[width=1\textwidth]{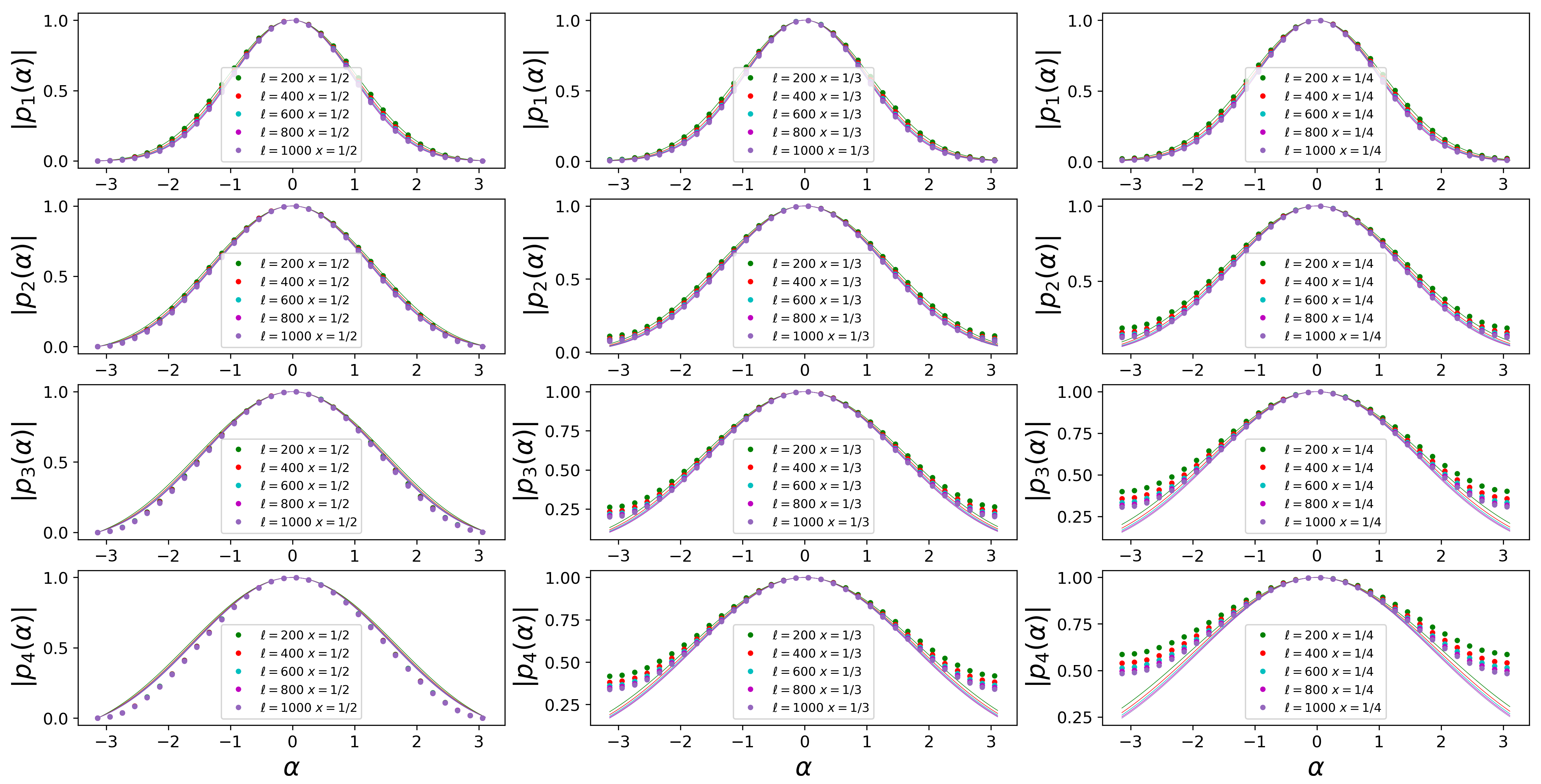}
\caption{Numerical data for the generating function $p_n(\alpha)$ for the particle-hole excitation in the XX chain (symbols). 
The full lines are the CFT predictions for the derivative operator, Eq. \eqref{fnpns} with \eqref{pol}; for the ground-state variance we use the exact results from 
Ref. \cite{Ric}. 
Here we consider $n=1,2,3,4$ (from top to bottom) and $x=1/2,1/3,1/4$ (from left to right).
We report several values of $\ell$ up to 1000; also the CFT prediction does depend on $\ell$ through the variance of the ground state
(the curves at different $\ell$ follows the same color code as the data).  
Again, the agreement is very good for small $\alpha$, but it worsens as $\alpha$ gets closer to $\pm\pi$ 
and as $n$ gets larger.
}
\label{pna}
\end{figure}

We can finally discuss the generating functions $p_n(\alpha,x)$ themselves. 
Although $p_n(\alpha)$ is just the product of $f_n(\alpha,x)$ and the ground state distribution $p_n^{GS}(\alpha,x)$, 
it is still worth to compare the numerical data with the CFT for a twofold reason:
(i) The CFT prediction for $p_n(\alpha,x)$ displays an explicit dependence on $\ell$ through $p_n^{GS}(\alpha,x)$ and in particular through its variance;
(ii) Since $p_n(\alpha,x)$ decays as a Gaussian as $\alpha$ moves away from $0$, the large deviations observed for $f_n(\alpha,x)$ in Fig. \ref{fna} may get 
suppressed by multiplying it with the ground-state Gaussian distribution.  
The data for $p_n(\alpha,x)$ are reported in Fig. \ref{pna} for $n=1,2,3,4$. We clearly observe that the matching of the CFT predictions and 
numerical data is improved compared to the ones for $f_n(\alpha,x)$ as a consequence of the multiplication by the Gaussian.
Anyhow, for large $n$ and $\alpha$ clear deviations are still evident, as expected.

\subsubsection{Symmetry resolved entropies}

\begin{figure}[!t]
\includegraphics[width=1\textwidth]{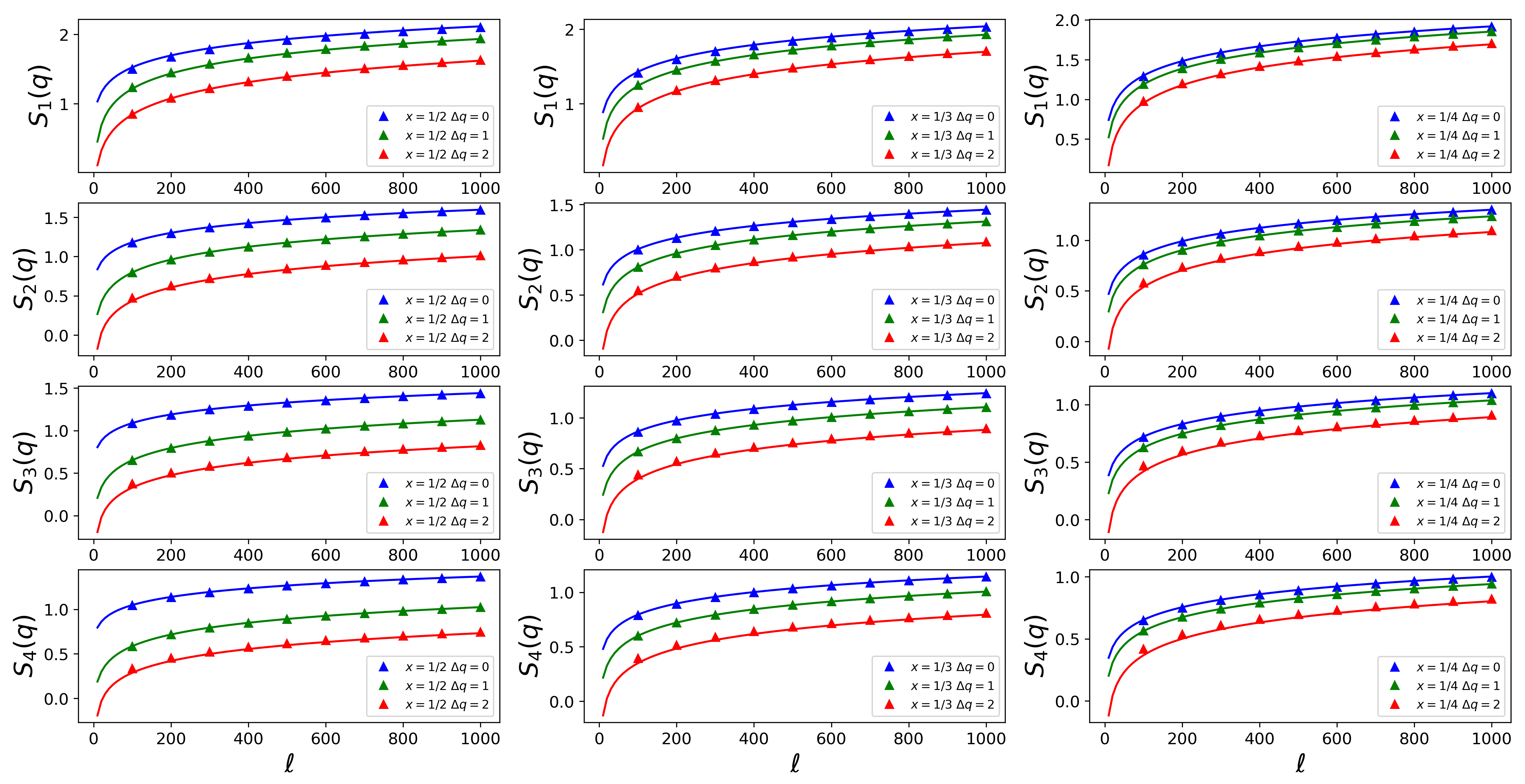}
\caption{Numerical data for the symmetry resolved entanglement  entropies  $S_n(q)$ for the particle-hole excitation in the XX chain (symbols) as function of $\ell$. 
The full lines are the CFT predictions for the derivative operator. 
For the ground-state $S_n(q)$ we use the exact results from Ref. \cite{Ric}. 
Here we consider $n=1,2,3,4$ (from top to bottom) and $x=1/2,1/3,1/4$ (from left to right).
In each panel we report three curves with $\Delta q \equiv q-\bar q=0,1,2$. 
}
\label{FSnq}
\end{figure}

In this section we compute the symmetry resolved entanglement. As a first step we must compute the probability distributions $p_n(q)$.
The Fourier transform of \eqref{expr} gives (up to the normalisation factor)  the probabilities $p_n(q)$. 
An efficient way to implement such Fourier transform is to write $\text{tr}(\rho_A^n e^{i\alpha Q_A})$ explicitly as a polynomial in $e^{i\alpha}$: the contribution 
of the sector with $Q_A = q$ is given by the coefficient of the $e^{i\alpha q}$ term. For $n \neq 1$, one can obtain $S_n(q)$ 
using Eq. \eqref{eq11} and computing separately $S_n$, $p_n(q)$ and $p(q)$. 
The case $n=1$ is singular, nevertheless the Fourier transform of the following expression
\begin{multline} \label{SvNnum}
-\text{tr}(\rho_A\log \rho_A e^{i\alpha Q_A}) = -\partial_n\text{tr}(\rho_A^n e^{i\alpha Q_A})|_{n=1}=\\
\sum_k (-\nu_k\log \nu_k e^{i\alpha} -(1-\nu_k)\log (1-\nu_k))\prod_{k'\neq k}(\nu_{k'}e^{i\alpha}+(1-\nu_{k'})),
\end{multline}
provides an efficient way to compute $Sp(q) - \partial_n p_n(q)|_{n=1}$ and therefore $S$ (using Eq.~\eqref{eq1}). 

We do not discuss here the intermediate results for $p_n(q)$ since they depend on too many variables ($x,q,n, \ell$) which are difficult to put together in a 
clear plot. Hence we just discuss the symmetry resolved entropies. 
The numerical data for $S_n(q)$ as function of $\ell$ for few values of $x=1/2,1/3,1/4$ and $n=1,2,3,4$ are reported in Fig. \ref{FSnq}. 
We focus on the most probable values of $q$ with $\Delta q \equiv q-\bar q=0,1,2$. 
In the plots, the CFT predictions for $n=2,3,4$ are obtained as numerical Fourier transform of the exact $p_n(\alpha)$ obtained in Fig. \ref{pna}. 
This is not possible for $n=1$; in this case we employ the quadratic approximation \eqref{q_app} which can be analytically continued; 
within this approximation the von Neumann entropy is just the limit for $n\to1$ of Eq. \eqref{Snq}. 
We see that the agreement of the CFT predictions with the numerical data in Fig. \ref{FSnq} is very good for all the values of the parameters 
we considered.

\section{Discussion \& Outlook} \label{sec:final}

In this manuscript, we fully characterised the symmetry resolved entanglement of excited states of two-dimensional CFTs generated by primary operators.
The first main result is Eq. \eqref{main} for the scaling function of the charged moments in a general theory with  a $U(1)$ symmetry.
This expression has been then explicitly evaluated in the free compact boson theory (Luttinger liquid) for the vertex and the derivative operators.
While for the vertex the outcome is trivial, for the derivative the final result \eqref{res2} is highly non-obvious and with many interesting physical features
discussed in the text. 
In particular, we found that all  the differences of cumulants between the excited states and ground state
are {\it universal}, i.e. do not depend on the microscopic details and are solely fixed by conformal invariance. 
From the Fourier transform of these charged moments, we extract the symmetry resolved entanglement R\'enyi entropies, stressing 
their universal aspects, such as a term breaking entanglement equipartition at order $(\log\ell)^{-2}$ within CFT. 
We tested our analytic predictions against exact numerical calculations in the XX spin chain, finding a perfect agreement. 
Incidentally, our results for $n=1$  are the full counting statistics (FCS) of the charge operator within an interval in these excited states of the CFT.
To the best of our knowledge, also these findings for the FCS and for the related probability are new  and generalise  the results for the 
ground state \cite{bss-07,aem-08}. 

While here we focused on low-lying excited states induced by primary operators in CFT, the same method can be applied to any excited state. 
In this direction, it would be interesting to study excited states generated, for instance, by descendent operators. 
Results are already available for the total entanglement and R\'enyi entropies \cite{palmai, top-16,cghw-15}, and therefore they can be generalised 
to the symmetry resolved ones. 
However, working out the explicit expressions, as usual, may become quite cumbersome.
Another natural development should be to study the symmetry resolved entanglement for excited states that are in the middle of the many-body spectrum.
These are characterised by a volume law \cite{afc-09,ly-15,aef-14,mbsa-14} and their physics is closely related to (generalised) 
eigenstate thermalisation hypothesis \cite{dls-13,alba-15,alba-2016,alba-2018}.   

Our results also can be extended to work out other entanglement-related quantities in their symmetry resolved fashion. For example, a natural extension would be to consider the relative entropy or the trace distance (both measuring distances between density matrices) by combining the results in \cite{Gsela} with those in Refs.~\cite{lashkari, ruggiero-relative, TD, TD-long}, where similar replica tricks for such quantities have been introduced.

Different symmetry resolved entanglement measures, such as logarithmic negativity in the ground state, have been worked out as well \cite{GS-neg}, 
again starting from the proper replica trick \cite{cct-neg}. 
In principle, also for the negativity and its symmetry resolution, one might wonder how to adapt the framework to excited states. 
However, in this case, some technical issues arise: one should deal with correlation functions on more complicated Riemann surfaces, which cannot be mapped 
to the complex plane. 
Still, the problem could in principle be approached through the techniques of Ref.~\cite{estienne} or approximation methods, as the ones of 
Refs.~\cite{GR, ruggiero-zamo} based on operator product expansion and recursive formulas for conformal blocks.

\section*{Acknowledgments}
PC acknowledges support from ERC under Consolidator grant number 771536 (NEMO).

\begin{appendix}

\section{Characteristic polynomial $P_M (\lambda)$} 
\label{appendixA}

In order to match the notations of \cite{det}, we rewrite the matrix $M$ in \eqref{matrixM} as
\begin{equation}
M = \frac{1}{2}\begin{pmatrix} A & B \\ -B^T & A\end{pmatrix}.
\end{equation}
with
\begin{equation}
A_{ij} = \begin{cases} 0  &\quad  \text{if}  \ i=j, \\ \frac{1}{\sin(\pi \frac{j-i}{n})} &\quad \text{otherwise},\end{cases} \qquad
B_{ij} = \frac{1}{\sin(\pi\frac{j-i-x}{n})}.
\end{equation}
The characteristic polynomial now takes the form $P_M(\lambda) = \text{det}(M-\lambda) = \frac{1}{4^n}\text{det}[(A-2\lambda)^2+B^T B]$.\\
A direct calculation shows that $B^TB = \alpha \cdot \mathbb{I}$ with $\alpha = \frac{n^2}{\sin^2 \pi x}$.
Moreover,  the expansion of the Newton polynomial gives
\begin{equation}
\tr[(A-2\lambda)^{2k}] = \sum_{j=0}^{2k}(-1)^j \binom{2k}{j}\text{tr}[A^j](2\lambda)^{2k-j},
\end{equation}
while the traces of the powers of $A$ are
\begin{equation}
\tr[A^{2k}] = 2(-1)^k\sum_{p=1}^{\lfloor n/2 \rfloor}(n-2p+1)^{2k}.
\end{equation}
With simple manipulations the previous expression becomes
\begin{multline}
P_M(\lambda) = \frac{\alpha^n}{4^{n}}\det(1 +\frac{(A-\lambda)^2}{\alpha}) = \left( \frac{\alpha}{4} \right)^n \exp \left[ \sum_{k=1}^\infty \frac{(-1)^{k+1}}{k\alpha^k} \right]=\\
= \left( \frac{\alpha}{4} \right)^n \exp [2 \sum_{k=1}^\infty \frac{(-1)^{k+1}}{k\alpha^k}\sum_{p=1}^{[n/2]}\sum_{j=0}^k\binom{2k}{2j} (-1)^j (2\lambda)^{2(k-j)} (n-2p+1)^{2j}].
\end{multline}
It is also possible to explicitly perform the sum in $j$ exploiting the following trick
\begin{equation}
\sum_{j=0}^k\binom{2k}{2j}x^{2j}(iy)^{2k-2j} = \frac{1}{2}[(x+iy)^{2k} + (x-iy)^{2k}],
\end{equation}
to eventually get
\begin{equation}
P_M(\lambda) = \left( \frac{\alpha}{4} \right)^n \left( \prod_{p=1}^{[n/2]}(1-\frac{(n-2p+1-2i\lambda)^2}{\alpha}) \times c.c. \right).
\end{equation}
In order to simplify the expression above, we recall some properties of the function $\Gamma(z)$
\begin{equation}
\frac{\Gamma(z+n+1)}{\Gamma(z+1)} = \prod_{k=1}^n(z+k), \qquad 
\Gamma(z)\Gamma(1-z)=\frac{\pi}{\sin(\pi z)}.
\end{equation}
Defining
\begin{equation}
z_0(\lambda)  = \frac{1}{2}\left( \frac{n}{\sin(\pi x)} -n -1 \right)+i\lambda,
\end{equation}
one finally obtains
\begin{equation} \label{PMfinal}
P_M(\lambda) = \frac{\Gamma(1+z_0(\lambda)+n)}{\Gamma(1+z_0(\lambda))} \frac{\Gamma(1+\bar{z}_0(\lambda)+n)}{\Gamma(1+\bar z_0(\lambda))}.
\end{equation}

\end{appendix}



\begin{thebibliography}{99}
\bibitem{hlw-94}
C. Holzhey, F. Larsen, and F. Wilczek, {\it Geometric and renormalized entropy in conformal field theory}, 
\href{http://dx.doi.org/10.1016/0550-3213(94)90402-2}{Nucl. Phys. B {\bf 424}, 443 (1994)}.

\bibitem{vlrk-03}
G. Vidal, J. I. Latorre, E. Rico, and A. Kitaev, {\it Entanglement in quantum critical phenomena}, 
\href{http://dx.doi.org/10.1103/PhysRevLett.90.227902}{Phys. Rev. Lett. {\bf 90}, 227902 (2003)};\\
J. I. Latorre, E. Rico, and G. Vidal,
{\it Ground state entanglement in quantum spin chains},
\href{https://arxiv.org/abs/quant-ph/0304098}{Quant. Inf. Comp. {\bf 4}, 048 (2004)}.

\bibitem{cc-04}
P. Calabrese and J. Cardy, {\it Entanglement entropy and quantum field theory}, 
\href{http://dx.doi.org/10.1088/1742-5468/2004/06/P06002}{J.  Stat. Mech. P06002 (2004)}.

\bibitem{cc-09}
P. Calabrese and J. Cardy, {\it Entanglement entropy and conformal field theory}, 
\href{http://dx.doi.org/10.1088/1751-8113/42/50/504005}{J. Phys. A {\bf 42}, 504005 (2009)}.


\bibitem{lihaldane}
H. Li and F. D. M. Haldane, \emph{Entanglement Spectrum as a Generalization of Entanglement Entropy: Identification of Topological Order in Non-Abelian Fractional Quantum Hall Effect States}, 
\href{http://dx.doi.org/10.1103/PhysRevLett.101.010504}{Phys. Rev. Lett. {\bf 101},  010504  (2008).}


\bibitem{amico-2008} 
L. Amico, R. Fazio, A. Osterloh, and V. Vedral, \textit{Entanglement in many-body systems},
\href{http://dx.doi.org/10.1103/RevModPhys.80.517}{Rev. Mod. Phys. {\bf 80}, 517 (2008).}


\bibitem{calabrese-2009} 
P.~Calabrese, J.~Cardy, and B.~Doyon Eds, \emph{Entanglement entropy in extended quantum systems}, 
\href{http://dx.doi.org/10.1088/1751-8121/42/50/500301}{J. Phys. A {\bf 42}, 500301 (2009).}

\bibitem{eisert-2010}
J.~Eisert, M.~Cramer, and M.~B.~Plenio, \emph{Area laws for the entanglement entropy}, 
\href{http://dx.doi.org/10.1103/RevModPhys.82.277}{Rev. Mod. Phys. {\bf 82}, 277 (2010).}

\bibitem{rev-lafl}
N. Laflorencie, {\it Quantum entanglement in condensed matter systems}, 
\href{http://dx.doi.org/10.1016/j.physrep.2016.06.008}{Phys. Rep. {\bf 643}, 1 (2016)}.


\bibitem{Gsela}
M. Goldstein and E. Sela, 
{\it Symmetry-Resolved Entanglement in Many-Body Systems}, 
\href{https://journals.aps.org/prl/abstract/10.1103/PhysRevLett.120.200602}{Phys. Rev. Lett. {\bf 120}, 200602 (2018)}.

\bibitem{GS-neg}
M. Goldstein and E. Sela, 
{\it Imbalance Entanglement: Symmetry Decomposition of Negativity}, 
\href{https://doi.org/10.1103/PhysRevA.98.032302}{Phys. Rev. A {\bf 98}, 032302 (2018)}.

\bibitem{equi-sierra}
J. C. Xavier, F. C. Alcaraz, and G. Sierra, 
{\it Equipartition of the entanglement entropy}, 
\href{https://journals.aps.org/prb/abstract/10.1103/PhysRevB.98.041106}{Phys. Rev. B {\bf 98}, 041106 (2018)}.

\bibitem{sara-gapped}
S. Murciano, G. Di Giulio, and P. Calabrese,
{\it Symmetry resolved entanglement in gapped integrable systems: a corner transfer matrix approach},
\href{https://arxiv.org/abs/1911.09588}{arXiv:1911.09588}.

\bibitem{Ric}
R. Bonsignori, P. Ruggiero, and P. Calabrese, \textit{Symmetry resolved entanglement in free fermionic systems},
\href{https://doi.org/10.1088/1751-8121/ab4b77}{J. Phys. A  \textbf{52}, 475302 (2019)}.


\bibitem{Gold}
S. Fraenkel and M. Goldstein,
\textit{Symmetry resolved entanglement: Exact results in 1D and beyond},
\href{https://arxiv.org/abs/1910.08459}{arXiv:1910.08459}.

\bibitem{lr-14}
N. Laflorencie and S. Rachel, {\it Spin-resolved entanglement spectroscopy of critical spin chains and Luttinger liquids}, 
\href{https://doi.org/10.1088/1742-5468/2014/11/P11013}{J. Stat. Mech. P11013 (2014)}.

\bibitem{sara-fcs}
P. Calabrese, M. Collura, G. Di Giulio, and S. Murciano, 
{\it Full counting statistics in the gapped XXZ spin chain},
\href{https://arxiv.org/abs/2002.04367}{arXiv:2002.04367}.


\bibitem{ryu-symmres}
M. T. Tan and S. Ryu,
{\it Particle Number Fluctuations, Renyi and Symmetry-resolved Entanglement Entropy in Two-dimensional Fermi Gas from Multi-dimensional Bosonization},
\href{https://arxiv.org/abs/1911.01451}{arXiv:1911.01451}.

\bibitem{mrc-20}
S. Murciano, P. Ruggiero, and P. Calabrese, to appear.

\bibitem{fg-19}
N. Feldman and M. Goldstein,
{\it Dynamics of Charge-Resolved Entanglement after a Local Quench},
\href{http://dx.doi.org/10.1103/PhysRevB.100.235146}{Phys. Rev. B {\bf 100}, 235146 (2019)}.

\bibitem{exp-lukin}
A. Lukin, M. Rispoli, R. Schittko, M. E. Tai, A. M. Kaufman, S. Choi, V. Khemani, J. Leonard, and M. Greiner,
{\it Probing entanglement in a many-body localized system},
\href{https://science.sciencemag.org/content/364/6437/256/tab-figures-data}{Science {\bf 364}, 6437 (2019)}.

\bibitem{excSierra}
F. C. Alcaraz, M. Ibanez Berganza, and G. Sierra, {\it Entanglement of Low-Energy Excitations in Conformal Field Theory},
\href{http://dx.doi.org/10.1103/PhysRevLett.106.201601}{Phys. Rev. Lett. {\bf 106}, 201601(2011)}.

\bibitem{excSierra2}
M. Ibanez Berganza, F. C. Alcaraz, and G. Sierra, \emph{Entanglement of excited states in critical spin chains}, 
\href{http://dx.doi.org/10.1088/1742-5468/2012/01/P01016}{J. Stat. Mech.  P01016 (2012)}.

\bibitem{pagialle}
P. Di Francesco, P. Mathieu, and D. Senechal, {\it Conformal Field Theory} (Springer-Verlag, New York, 1997).

\bibitem{giam-b}
  T. Giamarchi, {\it Quantum physics in one dimension}, Clarendon press (2003).

\bibitem{elc-13}
F. H. L. Essler, A. M. L\"auchli, and P. Calabrese, {\it Shell-Filling Effect in the Entanglement Entropies of Spinful Fermions},
\href{http://dx.doi.org/10.1103/PhysRevLett.110.115701}{Phys. Rev. Lett. {\bf 110}, 115701 (2013)}.



\bibitem{det}
P. Calabrese, F. Essler, and A. L\"auchli, \emph{Entanglement entropies of the quarter filled Hubbard model}, 
\href{http://dx.doi.org/10.1088/1742-5468/2014/09/P09025}{J. Stat. Mech. (2014) P09025}.

\bibitem{palmai}
T. Palmai,
{\it Excited state entanglement in one dimensional quantum critical systems: Extensivity and the role of microscopic details},
\href{http://dx.doi.org/10.1103/PhysRevB.90.161404}{Phys. Rev. B {\bf 90}, 161404 (2014)}.

\bibitem{top-16}
L. Taddia, F. Ortolani, and T. Palmai,
{\it Renyi entanglement entropies of descendant states in critical systems with boundaries: conformal field theory and spin chains},
\href{http://dx.doi.org/10.1088/1742-5468/2016/09/093104}{J. Stat. Mech. (2016) 093104}.

\bibitem{cghw-15}
B. Chen, W.-Z. Guo, S. He, and Ji. Wu, {\it Entanglement entropy for descendent local operators in 2D CFTs},
\href{https://doi.org/10.1007/JHEP10(2015)173}{JHEP {\bf 10},  173 (2015)}.

\bibitem{wv-03}
H. M. Wiseman and J. A. Vaccaro, {\it Entanglement of Indistinguishable Particles Shared between Two Parties}, 
\href{https://doi.org/10.1103/PhysRevLett.91.097902}{Phys. Rev. Lett. {\bf 91}, 097902 (2003)}.

\bibitem{bhd-18}
H. Barghathi, C. M. Herdman, and A. Del Maestro, {\it R\'enyi Generalization of the Accessible Entanglement Entropy}, 
\href{https://doi.org/10.1103/PhysRevLett.121.150501}{Phys. Rev. Lett. {\bf 121}, 150501 (2018)}.

\bibitem{bcd-19}
H. Barghathi, E. Casiano-Diaz, and A. Del Maestro, {\it Operationally accessible entanglement of one dimensional spinless fermions},
\href{https://doi.org/10.1103/PhysRevA.100.022324}{Phys. Rev. A {\bf 100}, 022324 (2019)}.

\bibitem{kusf-20}
M. Kiefer-Emmanouilidis, R. Unanyan, J. Sirker, and M. Fleischhauer, {\it Bounds on the entanglement entropy by the number entropy in non-interacting fermionic systems},
\href{https://arxiv.org/pdf/2003.03112.pdf}{ArXiv:2003.03112}.

\bibitem{Laf}
H. F. Song, S. Rachel, C. Flindt, I. Klich, N. Laflorencie, and K. Le Hur. \textit{Bipartite Fluctuations as a Probe of Many-Body Entanglement},
\href{https://journals.aps.org/prb/abstract/10.1103/PhysRevB.85.035409}{Phys. Rev. B \textbf{85}, 035409 (2012)}.

\bibitem{Min}
P.~Calabrese, M.~Mintchev, and E.~Vicari, {\it Exact relations between particle fluctuations and entanglement in Fermi gases}, 
 \href{http://dx.doi.org/10.1209/0295-5075/98/20003}{EPL {\bf {98}}, 20003 (2012)}.

\bibitem{peschel2001}
M. C. Chung and I. Peschel, \emph{Density-matrix spectra of solvable fermionic systems}, 
\href{http://dx.doi.org/10.1103/PhysRevB.64.064412}{Phys. Rev. {\bf B 64}, 064412 (2001)}.

\bibitem{peschel2003}
I. Peschel, {\it Calculation of reduced density matrices from correlation functions},
\href{http://dx.doi.org/10.1088/0305-4470/36/14/101}{J. Phys. A {\bf 36}, L205 (2003)}.

\bibitem{pe-09} 
I. Peschel and V. Eisler, {\it Reduced density matrices and entanglement entropy in free lattice models},
\href{http://dx.doi.org/10.1088/1751-8113/42/50/504003}{J. Phys. A {\bf 42}, 504003 (2009)}.

\bibitem{sach-book}
S. Sachdev, \emph{Quantum Phase Transitions}, Cambridge University Press (2001).

\bibitem{atc-10}
V. Alba, L. Tagliacozzo, and P. Calabrese, 
{\it Entanglement entropy of two disjoint blocks in critical Ising models},
\href{http://dx.doi.org/10.1103/PhysRevB.81.060411}{Phys. Rev. B {\bf 81}, 060411 (2010)}.
 
 \bibitem{ip-10}
F. Igloi and I. Peschel, 
 {\it On reduced density matrices for disjoint subsystems}, 
\href{http://dx.doi.org/10.1209/0295-5075/89/40001}{EPL {\bf 89} 40001 (2010).}


\bibitem{Inhom}
S. Murciano, P. Ruggiero, and P. Calabrese,
{\it Entanglement and relative entropies for low-lying excited states in inhomogeneous one-dimensional quantum systems},
 \href{http://dx.doi.org/10.1088/1742-5468/ab00ec}{J. Stat. Mech. (2019) 034001}.
 

\bibitem{sublead}
J.~Cardy and P.~Calabrese, {\it Unusual Corrections to the Scaling in Entanglement Entropy}, 
\href{http://dx.doi.org/10.1088/1742-5468/2010/04/P04023}{J. Stat. Mech. (2010) P04023}.


\bibitem{ot-15}
K. Ohmori and Y. Tachikawa, {\it Physics at the entangling surface},
\href{http://dx.doi.org/10.1088/1742-5468/2015/04/P04010}{J. Stat. Mech. P04010 (2015)}. 



\bibitem{parity2}
P.~Calabrese and F.~H.~L. Essler,
{\it Universal corrections to scaling for block entanglement in spin-1/2 XX chains}, 
\href{http://dx.doi.org/10.1088/1742-5468/2010/08/P08029}{J. Stat. Mech. P08029 (2010)}.

\bibitem{Fagotti2010cc}
M.~Fagotti and P.~Calabrese,
{\it Universal parity effects in the entanglement entropy of XX chains with open boundary conditions}, 
\href{http://dx.doi.org/10.1088/1742-5468/2011/01/P01017}{J. Stat. Mech. P01017 (2011)}.




\bibitem{parity}
P.~Calabrese, M.~Campostrini, F.~H.~L. Essler, and B.~Nienhuis,
{\it Parity effects in the scaling of block entanglement in gapless spin chains}, 
\href{http://dx.doi.org/10.1103/PhysRevLett.104.095701}{Phys. Rev. Lett {\bf 104}, 095701 (2010)}.


\bibitem{correzioni}
P.~{Calabrese}, J.~Cardy, and I~Peschel,
{\it Corrections to scaling for block entanglement in massive spin-chains},
\href{http://dx.doi.org/10.1088/1742-5468/2010/09/P09003}{J. Stat. Mech. P09003 (2010)}.

\bibitem{overlap1}
P.~Calabrese, M.~Mintchev, and E.~Vicari,
{\it Entanglement Entropy of One-Dimensional Gases},  
\href{https://doi.org/10.1103/PhysRevLett.107.020601}{Phys. Rev. Lett. {\bf 107}, 020601 (2011)}.


\bibitem{mint4}
P.~Calabrese, M.~Mintchev, and E.~Vicari,
{\it The entanglement entropy of one-dimensional systems in continuous and homogeneous space}, 
\href{http://dx.doi.org/10.1088/1742-5468/2011/09/P09028}{J. Stat. Mech. P09028 (2011)}. 


\bibitem{num}
J. C. Xavier and F. C. Alcaraz,  {\it Renyi entropy and parity oscillations of anisotropic spin-s Heisenberg chains in a magnetic field},
\href{http://dx.doi.org/10.1103/PhysRevB.83.214425}{Phys. Rev. B {\bf 83},  214425 (2011)};\\
M. Dalmonte, E. Ercolessi, and L. Taddia, {\it Estimating quasi-long-range order via Renyi entropies},
\href{http://dx.doi.org/10.1103/PhysRevB.84.085110}{Phys. Rev. B {\bf  84}, 085110 (2011)}; \\
M. Dalmonte, E. Ercolessi, and L. Taddia, {\it Critical properties and Renyi entropies of the spin-3/2 XXZ chain}, 
\href{http://dx.doi.org/10.1103/PhysRevB.85.165112}{Phys. Rev. B {\bf 85}, 165112 (2012)}.


\bibitem{sublead1}
L.~Cevolani, {\it Unusual Corrections to the Scaling of the Entanglement Entropy of the Excited states in Conformal Field Theory},
\href{https://arxiv.org/abs/1601.01709}{arXiv:1601.01709}.


\bibitem{Caputa1}
P. Caputa, M. Nozaki, and T. Numasaw,  {\it Charged Entanglement Entropy of Local Operators},
\href{https://arxiv.org/abs/1512.08132}{arXiv:1512.08132}.

\bibitem{Caputa2}
P. Caputa, G. Mandal, and R. Sinha,  {\it Dynamical entanglement entropy with angular momentum and U(1) charge}, \href{https://arxiv.org/abs/1306.4974}{arXiv:1512.08132}.

\bibitem{CFH}
H. Casini, C. D. Fosco, and M. Huerta, 
{\it Entanglement and alpha entropies for a massive Dirac field in two dimensions}, 
\href{https://iopscience.iop.org/article/10.1088/1742-5468/2005/07/P07007}{J. Stat. Mech. (2005) P07007}.

\bibitem{ch-rev}
H. Casini and M. Huerta, {\it Entanglement entropy in free quantum field theory},
\href{https://doi.org/10.1088/1751-8113/42/50/504007}{J. Phys. A {\bf 42}, 504007 (2009)}.

\bibitem{d-16}
J. S. Dowker,  {\it Conformal weights of charged R\'enyi entropy twist operators for free scalar fields in arbitrary dimensions},
\href{https://doi.org/10.1088/1751-8113/49/14/145401}{J. Phys. A {\bf 49}, 145401 (2016)};\\
J. S. Dowker,  {\it Charged R\'enyi entropies for free scalar fields},
\href{https://doi.org/10.1088/1751-8121/aa6178}{J. Phys. A {\bf 50}, 165401 (2017)}.

\bibitem{matsuura}
A. Belin, L.-Y. Hung, A. Maloney, S. Matsuura, R. C. Myers, and T. Sierens, 
{\it Holographic charged R\'enyi entropies}, 
\href{https://doi.org/10.1007/JHEP12(2013)059}{JHEP {\bf 12} (2013) 059}.

\bibitem{cnn-16}
P. Caputa, M. Nozaki, and T. Numasawa, {\it Charged Entanglement Entropy of Local Operators}, 
\href{https://dx.doi.org/10.1103/PhysRevD.93.105032}{Phys. Rev. D {\bf 93}, 105032 (2016)}.

\bibitem{ssr-17}
H. Shapourian, K. Shiozaki, and S. Ryu, {\it Partial time-reversal transformation and entanglement negativity in fermionic systems}, 
\href{https://doi.org/10.1103/PhysRevB.95.165101}{Phys. Rev. B {\bf 95}, 165101 (2017)}.

\bibitem{shapourian-19}
H. Shapourian, P. Ruggiero, S. Ryu, and P. Calabrese,
{\it Twisted and untwisted negativity spectrum of free fermions},
\href{https://arxiv.org/pdf/1906.04211.pdf}{arXiv:1906.04211}




\bibitem{bss-07}
M. Bortz, J. Sato, and M. Shiroishi M, {\it String correlation functions of the spin-1/2 Heisenberg XXZ chain},  
\href{https://doi.org/10.1088/1751-8113/40/16/001}{J. Phys. A {\bf 40}, 4253 (2007)}.

\bibitem{aem-08}
D. B. Abraham, F. H. L. Essler, and A. Maciolek, {\it Effective Forces Induced by a Fluctuating Interface: Exact Results}, 
\href{https://doi.org/10.1103/PhysRevLett.98.170602}{Phys. Rev. Lett. {\bf 98}, 170602 (2007)}.

 \bibitem{afc-09}
V. Alba, M. Fagotti, and P. Calabrese, {\it Entanglement entropy of excited states},
\href{http://dx.doi.org/10.1088/1742-5468/2009/10/P10020}{J. Stat. Mech. (2009) P10020.}

\bibitem{mbsa-14}
J. Molter, T. Barthel, U Schollwock, and V. Alba, {\it Bound states and entanglement in the excited states of quantum spin chains},
\href{https://doi.org/10.1088/1742-5468/2014/10/P10029}{J. Stat. Mech. (2014) P10029}.

\bibitem{ly-15}
H.-H. Lai and K. Yang, {\it Entanglement entropy scaling laws and eigenstate typicality in free fermion systems}, 
\href{https://doi.org/10.1103/PhysRevB.91.081110}{Phys. Rev. B {\bf 91}, 081110 (2015)}.

\bibitem{aef-14}
F. Ares, J. G. Esteve, F. Falceto, and E. Sanchez-Burillo, {\it Excited state entanglement in homogeneous fermionic chains},
\href{https://doi.org/10.1088/1751-8113/47/24/245301}{J. Phys. A {\bf 47}, 245301 (2014)}.



\bibitem{dls-13}
J. M. Deutsch, H. Li, and A. Sharma, Microscopic origin of thermodynamic entropy in isolated systems, 
\href{http://dx.doi.org/10.1103/PhysRevE.87.042135}{Phys. Rev. E {\bf 87}, 042135 (2013)}.

\bibitem{alba-15}
V. Alba, {\it Eigenstate thermalization hypothesis (ETH) and integrability in quantum spin chains},
\href{http://dx.doi.org/10.1103/PhysRevB.91.155123}{Phys. Rev. B {\bf 91}, 155123 (2015)}



\bibitem{alba-2016}
V.~Alba and P.~Calabrese, Entanglement and thermodynamics after a quantum quench in integrable systems, 
\href{http://dx.doi.org/10.1073/pnas.1703516114}{PNAS {\bf 114}, 7947 (2017)}.

\bibitem{alba-2018}
V. Alba and P. Calabrese, Entanglement dynamics after quantum quenches in generic integrable systems,
\href{http://dx.doi.org/10.21468/SciPostPhys.4.3.017}{SciPost Phys. {\bf 4}, 017 (2018)}.

\bibitem{lashkari}
N. Lashkari, \emph{Relative entropies in conformal field theory},
\href{http://dx.doi.org/10.1103/PhysRevLett.113.051602}{Phys. Rev. Lett. {\bf 113}, 051602 (2014)};\\
N. Lashkari, \emph{Modular hamiltonian of excites states in conformal field theory}, 
\href{http://dx.doi.org/10.1103/PhysRevLett.117.041601}{Phys. Rev. Lett. {\bf 117}, 041601 (2016).}

\bibitem{ruggiero-relative}
P.~Ruggiero and P.~Calabrese,
{\it Relative Entanglement Entropies in 1+1-dimensional conformal field  theories},
\href{http://dx.doi.org/10.1007/JHEP02(2017)039}{JHEP \textbf{02}, 039 (2017)}.

\bibitem{TD}
J. Zhang, P. Ruggiero, and P. Calabrese,
\emph{Subsystem Trace Distance in Quantum Field Theory},
\href{https://link.aps.org/doi/10.1103/PhysRevLett.122.141602}{Phys. Rev. Lett. {\bf 122}, 141602}.

\bibitem{TD-long}
P. Zhang, P. Ruggiero, and P. Calabrese, 
\emph{Subsystem trace distance in low-lying states of (1 + 1)-dimensional conformal field theories},
\href{https://doi.org/10.1007/JHEP10(2019)181}{JHEP {\bf 10} (2019) 181}.

\bibitem{cct-neg}
P. Calabrese, J. Cardy, and E. Tonni, 
{\it Entanglement negativity in quantum field theory}, 
\href{http://dx.doi.org/10.1103/PhysRevLett.109.130502}{Phys. Rev. Lett. {\bf 109}, 130502 (2012)};\\
P. Calabrese, J. Cardy, and E. Tonni, 
{\it Entanglement negativity in extended systems: a quantum field theory approach}, 
\href{http://dx.doi.org/10.1088/1742-5468/2013/02/P02008}{J. Stat. Mech.  P02008 (2013)};\\
P.~Calabrese, L.~Tagliacozzo, and E.~Tonni, {\it Entanglement negativity in the critical Ising chain}, 
\href{https://doi.org/10.1088/1742-5468/2013/05/P05002}{J. Stat. Mech.  P05002 (2013)}. 

\bibitem{estienne}
T. Dupic, B. Estienne, and Y. Ikhlef, {\it Entanglement entropies of minimal models from null-vectors},
\href{https://scipost.org/10.21468/SciPostPhys.4.6.031}{SciPost Phys. {\bf 4}, 031 (2018)}.

\bibitem{GR}
M.~A.~Rajabpour and  F.~Gliozzi, \emph{Entanglement entropy of two disjoint intervals from fusion algebra of twist fields},
\href{http://iopscience.iop.org/article/10.1088/1742-5468/2012/02/P02016/meta}{J. Stat. Mech. P02016  (2012).}

\bibitem{ruggiero-zamo}
P. Ruggiero, E. Tonni, and P. Calabrese,
\emph{Entanglement entropy of two disjoint intervals and the recursion formula for conformal blocks},
\href{https://iopscience.iop.org/article/10.1088/1742-5468/aae5a8}{J. Stat. Mech. (2018) 113101}.


\end{thebibliography}
\end{document}